\documentclass{article}

\usepackage{arxiv}

\usepackage[utf8]{inputenc} 
\usepackage[T1]{fontenc}
\PassOptionsToPackage{hyphens}{url}
\usepackage{hyperref}       
\usepackage{url}            
\usepackage{booktabs}       
\usepackage{amsfonts}       
\usepackage{nicefrac}       
\usepackage{microtype}      
\usepackage{lipsum}
\usepackage{graphicx}
\usepackage{algorithm,algpseudocode}
\usepackage{datatool}	

\algtext*{EndWhile}
\algtext*{EndIf}
\algtext*{EndFor}
\algtext*{EndProcedure}
\algtext*{EndFunction}
\algdef{SE}[DOWHILE]{DoWhile}{EndDoWhile}{\algorithmicdo}[1]{\algorithmicwhile\ #1}%
\algnewcommand\algorithmicforeach{\textbf{for each}}
\algdef{S}[FOR]{ForEach}[1]{\algorithmicforeach\ #1\ \algorithmicdo}
\algnewcommand\algorithmicand{\textbf{and}}
\algnewcommand\algorithmicnot{\textbf{not}}
\algnewcommand\algorithmicor{\textbf{or}}
\algnewcommand\algorithmicin{\textbf{in}}
\algnewcommand{\IIf}[1]{\State\algorithmicif\ #1\ \algorithmicthen}
\algnewcommand{\IElse}[1]{\State\algorithmicelse\ #1}
\algnewcommand{\EndIIf}{\unskip\ \algorithmicif}
\algnewcommand\Initialize{\item[\textbf{Initialize:}]}%
\algnewcommand{\LeftComment}[1]{\Statex \triangleright\ #1}

\title{Openbots}

\renewcommand{\subtitle}{An Empirical Study on Automated Programs in Social Media}

\author{
  Dennis Assenmacher \\
  Department of Information Systems and Statistics\\
  University of Münster\\
  Münster, 48149 \\
  \texttt{dennis.assenmacher@uni-muenster.de} \\
     \And
  Lena Adam \\
  Department of Information Systems and Statistics\\
  University of Münster\\
  Münster, 48149 \\
  \texttt{lena.adam@uni-muenster.de} \\
   \And
  Lena Frischlich \\
  Department of Communication\\
  University of Münster\\
  Münster, 48149 \\
  \texttt{lena.frischlich@uni-muenster.de} \\
     \And
Heike Trautmann \\
  Department of Information Systems and Statistics\\
  University of Münster\\
  Münster, 48149 \\
  \texttt{heike.trautmann@uni-muenster.de} \\
   \And
 Christian Grimme \\
  Department of Information Systems and Statistics\\
  University of Münster\\
  Münster, 48149 \\
  \texttt{christian.grimme@uni-muenster.de} \\
}

\begin{document}
\maketitle

\begin{abstract}
Social bots have recently gained attention in the context of  public opinion manipulation on social media platforms. While a lot of research effort has been put into the classification and detection of such ~(semi-)automated programs, it is still unclear how sophisticated those bots actually are, which platforms they target, and where they originate from. To answer these questions, we gathered repository data from open source collaboration platforms to identify the status-quo as well as trends of publicly available bot code. Our findings indicate that most of the code on collaboration platforms is of supportive nature and provides modules of automation instead of fully fledged social bot programs. Hence, the cost (in terms of additional programming effort) for building social bots with the goal of topic-specific manipulation is higher than assumed and that methods in context of machine- or deep-learning currently only play a minor role. However, our approach can be applied as multifaceted knowledge discovery framework to monitor trends in public bot code evolution to detect new developments and streams. 

\end{abstract}


\section{Introduction}
\label{sec:intro}

The emergence of the internet and the global success of social media has fundamentally changed how people search for information, how they are connected to each other, and how they communicate with each other \cite{Vorderer2015d}. Social media platforms like Facebook, Instagram, or Twitter as well as instant messengers like WhatsApp have become an established part of the modern media diet~\cite{Newman2017}. In 2018, a majority of Americans used Facebook and nearly a quarter was active on WhatsApp \cite{Smith2018}. In other countries, the proportion of WhatsApp users is even higher: about 70\% of the Germans use WhatsApp - often on a daily basis~\cite{Whatsbroadcast.com2018}. Social Media and Instant Messengers offer enhanced possibilities for interpersonal communication largely without geographical or temporal boundaries. Additionally, in some regions, social media is increasingly used as central communication and information channel, e.g., WhatsApp in India~\cite{StatistaIndia2016,Raina2018}. 

Not all communication partners in online-interactions are humans. 
Fully or semi-automatized user-accounts, so-called \emph{Social bots}, increasingly participate in online interactions. 
In contrast to other automated agents (such as web-crawlers or service bots), social bots are designed for one- or many-sided communication and the imitation of human online-behavior~\cite{Grimme2017,Frischlich2018}. The spectrum of (assumed or observed) types of social bots ranges from very simple bots that automate single elements of the communication process (e.g. liking or sharing), over partially human-steered accounts with automated elements (so-called hybrid bots, or  "Cyborgs", see~\cite{Chu2010}) to autonomously acting agents~\cite{Grimme2017} equipped with artificial intelligence and learning skills such as Microsofts' Zo (https://www.zo.ai/). Currently, social bots and their influence are especially discussed in the context of manipulation and disinformation~\cite{KollanyiUSEl2016,Ferrara2016,bessi2016social}. However, the detection of social bots remains a challenge~\cite{Grimme2018}, wherefore the actual number of social bots and also details of realization are unclear. Varol and colleagues~\cite{Varol2017} estimated that in 2017, a fraction of 9-15\% of the active twitter accounts were social bots, while platforms themselves report on millions of accounts~\cite{Roth2018}.

It is clear, that Social bots need a technical online infrastructure, which
which in general can be broadly understood as the combination of (a) a user-account on a social network site, a micro-blogging service, or an instant messenger; and (b) the technical preconditions for partial automation of the accounts-behavior via the accordant platform's API or proprietary mechanisms to interact with the website or app front-end of the account (this is virtually equivalent to remote controlling an app or browser). Additionally, the technical perspective includes the algorithmic realization of the account's behavior. While a basic social bot may only perform simple amplification activities (liking or sharing), more advanced bots may possibly replicate human behavior, post messages, or even directly interact with other accounts (humans or bots).
Apart from multiple often theoretical or anecdotal taxonomies, only little is known about the relative availability of these different types of social bots. From the technical perspective, the term "availability" implies three consecutive research questions to be posed: 

\begin{enumerate}
    \item \emph{What amount of ready-to-use code is publicly available to realize social bots?} This question addresses the general and free availability of source code in form of automation, access, and remote control frameworks, or ready-to-use code blocks to build social bots.
    \item \emph{What role does machine learning and artificial intelligence play for current off-the-shelf bot development?}. Considering the multiple taxonomies for social bots as well as public discussion on social bots, both (at least implicitly) assume the existence of what is often called "intelligent" bots. This question investigates whether a corresponding development is identifiable in open bot code.
    \item \emph{Consequently, what are the costs for creating social bots of different complexity levels (simple up to artificially intelligent)?} Depending on the availability of ready-to-use code for social bot development and based on the findings regarding available ingredients for constructing simple up to "intelligent" bots, we can roughly infer the effort, which is necessary to construct different types of social bots.
\end{enumerate}

This work strives to answer the questions by exploratory knowledge discovery using data from open-access social bots development projects, which is shared on different online-repositories, as an estimator. Online repository and collaboration infrastructures such as the 
platform \texttt{Github}\footnote{\url{https://github.com/}} have become a central hub of current software development processes~\cite{Zagalsky2015} and thus provide meaningful insights into the available knowledge in a certain programming area. 

Based on the insights gained from the worldwide available code for social bots, we propose a set of analytic methods that provides a broad picture to which extent social bots can be constructed from freely available code building blocks, frameworks, and technology snippets. This enables us to judge on the common stage of development of social bots "in the developer market" and can provide - on the long term - a way to detect new developments, threats, or trends in social bot evolution.

The paper is structured as follows: Section~\ref{sec:related_work} provides a literature overview on related work and the state of knowledge about social bot technology. Section~\ref{sec:data_sources} details the data source selection, relevant social media platforms, and data acquisition process. The analysis starts in Section~\ref{sec:descr_analysis} with a description of the acquired data and is continued in Section~\ref{sec:qual_analysis} with the data exploration and qualitative analysis of the social bot code descriptions using topic modelling via Latent Dirichlet Allocation (LDA)\cite{lda} and an innovative approach of stream clustering for texts~\cite{ER17Twitch}. Finally, in Section~\ref{sec:discussion}, we discuss our results and conclude our work.

\newpage
\section{Related Work}
\label{sec:related_work}

So far, research on social bots mainly addressed specific (sometimes learning- and thus data-based) approaches for the identification and observation of automated accounts in online media~\cite{Paradise2014,Cao2014,clark2016a,Ferrara2016,Varol2017,GG2018,Cornelissen2018ANT} or, discussed the influence of malicious social bots on the public debate~\cite{Fredheim2013,AYM2015,bessi2016social,Marechal2016} for instance via the distribution of hate or fake news~\cite{Frischlich2018,Shin2018,Vargo2018}.

Based on observations, detection of suspicious and clearly automated accounts, practice reports, and probably biased by public discussion, several social bot classifications of taxonomies have been proposed. As an early and rough classification Wooley~\cite{Woolley2016} distinguishes between classical bots for pure automation purposes and those who influence "public opinion or disrupt organizational communication". He defines such bots as "political bots". Hegelich~\cite{hegelich2016} follows a similar approach by defining two classes: "assistants" (also chat bots) and "social bots", where the latter are hidden actors in a political context. A more detailed categorisation is provided by Stieglitz and colleagues~\cite{Stieglitz2017DoSB}. They assume "imitation of human behaviour" and "intent" as two nominal scales, where the first discriminates low/none and high while the latter distinguishes malicious, neutral and benign. In this matrix, social bots imitate human behavior with malicious intent. Gorwa and Guilbeault~\cite{GG2018} address the "incredible breadth of terminology" used for (social) bots in literature. Focusing on the functions of bots, they suggest a more holistic typology: Crawlers and scrapers "working behind the scenes"; chat bots for human computer interaction using text; spam bots that spread advertisement, comments, or perform DDoS attacks in an automated and high frequency fashion; social bots, which the authors classify as web 2.0 versions of bots that often use APIs, however, are not clearly distinguishable in function from previously mentioned bots - except those, that they can act politically. In addition the authors also mention sockpuppets and trolls as well as cyborgs and hybrid accounts as partly automated profiles and thus special categories.
In an attempt to integrate the variety of perspectives, Grimme and colleagues~\cite{Grimme2017} propose a taxonomy that first distinguishes social and non-social bots (the latter corresponds to assistants mentioned by Hegelich~\cite{hegelich2016} or Gorwa and Guilbeault~\cite{GG2018}) and splits the class of social bots into three sub-classes: simple, advanced, and hypothetically "intelligent" social bots. For simple bots, the authors provide source code, which is able to perform very simple tasks like posting, sharing, or liking content. The second class is considered to imitate human behavior (in the sense of Stieglitz et al.~\cite{Stieglitz2017DoSB}) by acting in human speed, mimicking human inactivity, and simulating off-topic interests. Here, the authors report on experiments they performed with this kind of bots. Finally, they assume a possible third class of intelligent bots, that act on their own and virtually human-like, also in content production. However, as the authors emphasize, for this class no representative set of instances has yet been detected or published.

In fact, apart from an enormous set of online (gray) literature\footnote{e.g. BotWiki \url{https://botwiki.org/resources/twitterbots/} , Fredheim \url{http://quantifyingmemory.blogspot.co.uk/2013/06/putins-bots-part-one-bit-about-bots.html}, Grossmann \url{https://medium.freecodecamp.org/my-open-source-instagram-bot-got-me-2-500-real-followers-for-5-in-server-costs-e40491358340} } 
or technology documentations\footnote{e.g. \url{https://github.com/eggheads/eggdrop/}, \url{https://dev.botframework.com/}} 
, only very few scientific works back the postulated taxonomies or findings with explicit expertise on actual social bot code. Grimme et al.~\cite{Grimme2017,Grimme2018} are rare examples of this group. As such, the empirical evidence on available software to realize social bots as well as insight into the degree of development of such codes is very scarce. In a notable exception, in 2016, Kollanyi~\cite{Kollanyi2016} has examined the availability of open-source code for twitter bots on \texttt{Github}. He shows that the number of repositories providing twitter bot code has been steadily increasing since the launch of the platform, with the majority of repositories being provided by actors from the United State or Japan. In light of changing media system and the global success of platforms besides Twitter, it is, however, highly plausible that Twitter bots only form a small share of the overall social bots available. As such the study by Kollanyi --- albeit providing pioneering insights into the availability of open-source social bot code in general and their abilities more specifically --- might be  limited when it comes to understanding the availability of social bots more generally. 

The current study, however, builds on Kollanyi's approach and aims for a set of methodologically advanced tools for providing insights into (1) the availability of social bot code for the most successful social media platforms as well as instant messengers all over the world and (2) the skills these bots have implemented in their code related to the taxonomy of Grimme et al.~\cite{Grimme2017}.

\section{Data Sources}
\label{sec:data_sources}

\paragraph{Code Repositories:}
%
%
%
%

In the first step, a total of 54  active online code repositories were identified. As we were interested in the available knowledge about social bots, we focused on open platforms that allowed for collaboration and identification of specific codes via searchterms. 
A total of eight repositories fulfilled these criteria and allowed for (a) version control via Git or Appache subversion (SVN); (b) collaboration between users; (c) public access (that is they were researchable in the clear web);  and (d) searching for specific terms. 
For the analyses, we used Alexa global usage statistics to identify the five most relevant repositories. The Alexa rank, is a metric, which can be taken to evaluate the importance of a website\footnote{see: Alexa Rank: https://www.alexa.com/}. The metric combines calculations of internal homepage traffic such as page callings, and their development over time. Websites are ranked by their importance, where an Alexa score of "1", means that the website is most important. The ranking is seen critically in practice, since it is prone to get manipulated e.g. by click spamming and further does not differentiate between specific services or website purposes~\cite{alexa}. Nevertheless, it can be used to compare website services within the same topics of interest. Table \ref{Table_Top5} shows the final selection of relevant code repositories. 

\begin{table}\small
\setlength{\tabcolsep}{20pt}
\centering
\caption{Top 5 code repository hosting platforms}
\label{Table_Top5}
\begin{tabular}{@{}llr@{}}
\toprule
\textbf{Rank} & \textbf{Code Repository} & {\textbf{Alexa Rank (September 2018)}} \\ \midrule
1             & GitHub             & 72                                                  \\
2             & Sourceforge        & 351                                                 \\
3             & Bitbucket          & 945                                                 \\
4             & GitLab             & 2,819                                               \\
5             & Launchpad          & 7,529                                               \\
\bottomrule
\end{tabular}
\end{table}

\paragraph{Social Media Platforms:}
In order to describe the availability of different types of social bots, we focused on social media platforms, micro-blogging services, chat or Voice over IP-services, and instant messenger with the largest global reach. Reach was determined by a triangulation of (a) number of active users (see \cite{Statista2018}), (b) global traffic rank based on Alexa, and (c) downloads of the accordant application (e.g. via Google's Play Store or Apple's App Store). Where traffic information was missing (e.g. for the Instant Messengers) only applications with more than 500 Million downloads were included. As we focused on English-speaking repositories, the Chinese platforms were excluded from the key-word selection. 
Based on the criteria, mentioned above, the following  social media platforms were used within the data-acquisition process: \textit{Telegram, Twitter, Facebook, Reddit, Skype, Instagram, Youtube, Whatsapp, Linkedin, Tumblr, vKontakte, Snapchat and Pinterest}. 

\paragraph{Data Acquisition:}
Since the collaboration platforms are differently structured, it was not feasible to establish a common and comparable procedure for searching for specific bot programs. The largest platform, \texttt{Github}, offers a detailed search engine where explicit search criteria can be applied on different repository fields. Similar to \texttt{Github}, \texttt{Gitlab} also offers an API. However, \texttt{Gitlab} is more limited when it comes to the specification of additional search criteria. \texttt{Bitbucket} only provides a single web-search interface without any documentation. Therefore it is infeasible to track which fields are used within the final search query. Given the diverse capabilities of the collaboration platforms, when it comes to the formulation of searchterm queries, we decided to dismiss any field restriction at all. Additionally we selected the searchterms as generic as possible. Concretely we combined the name of each Social Platform with the term \texttt{bot} via a logical AND operator. For \texttt{Github}, \texttt{Gitlab} and \texttt{Bitbucket} a unique crawler was programmed that automatically gathered the repositories information for all searchterm combinations. While \texttt{Github} and \texttt{Gitlab} explicitly provide an external API for searching, \texttt{Bitbucket} is not easily accessible. Therefore we utilized \texttt{Scrapy}, a python-based web scraping framework, for collecting the relevant information. The remaining platforms, Sourceforge and Launchpad were manually queried via the provided web interface because of the low number of matching repositories for those platforms. The scraped information was persisted within Elasticsearch, a document-based search engine which performs well on textual data. 


To allow for time efficient crawling and avoid noise in the data set due to temporal developments during data collection, we specified the following limitations to our gathering process: First, we did not download the actual files (source code) of the repositories, since our analysis is mainly based on metadata. Secondly, we dismissed the history of individual commits (code contributions) on all repositories. Although these data may provide interesting insights, the amount of potential additional API requests would have been significantly increased. Instead, we limit our analysis to the first and last contribution. Due to the heterogeneous structure of the collaboration platforms, we defined a common intermediate schema for data representation. Although some platforms consist of fields (location attribute on \texttt{Github}) that are not present on other sources, we include these additional information sources in our analysis. This especially holds true for the \texttt{Github} platform which contains more than 90\% of all repositories. 

\section{Descriptive Analysis}
\label{sec:descr_analysis}
In total the data of 40.301 different code-repositories was gathered between April 2008 and October 2018. The largest number of repositories was provided by \texttt{Github}(38.600), followed by \texttt{Gitlab}(1293) and \texttt{Bitbucket}(408).
Despite its high Alexa score, only 25 repositories were found on Sourceforge for all searchtearm combinations. Moreover, 10 of these repositories were  maintained on \texttt{Github} in parallel. We explain this observation by the fact that Sourceforge is considered as one of the older collaboration platforms, with a lack of sophisticated functionality. Therefore, most developers probably decided to move to a different platform, which was able to fulfill their requirements. Also in 2013 and 2015 the platform was criticized for offering adware and bundled malware. As a result it was reported that users switched to other code-hosting platforms \cite{reutersgitlab}. For Launchpad, only 10 repositories were found. This is not a surprising result, since the platform is of small scale. In total the platform hosts only 13.000 repositories laying the focus on big, open source software projects such as MySQL, Inkscape or Unity.

\begin{figure}
\centering
\includegraphics[width=0.7\textwidth]{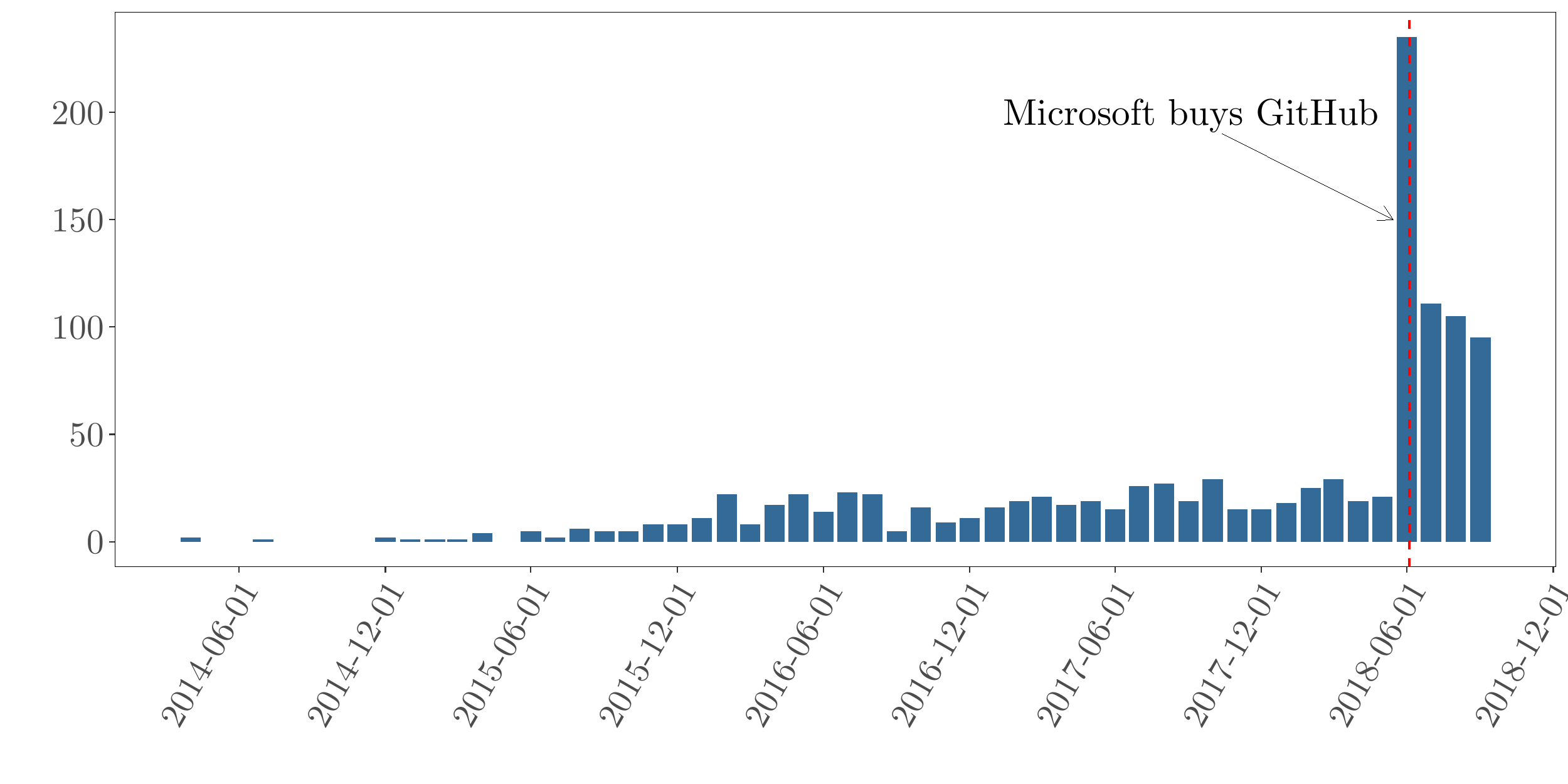}
\caption{Number of new \texttt{gitlab} repositories over time}
\label{fig:gitlab}
\end{figure}

The largest competitors of \texttt{Github}, namely \texttt{Gitlab} and \texttt{Bitbucket}, provide only a small fraction of the total number of bot repositories (4\%) and are thus considered as niche platforms. Furthermore, we are able to observe the impact of Microsoft's recent announcement of aquiring the \texttt{Github} platform for 7.5 billion US dollars \cite{microsoftgit}. While the average number of new repositories on the \texttt{Gitlab} platform per month was 13.97 before the announcement, it drastically increased to 234 repositories in June and 117 in July 2018 (see Figure \ref{fig:gitlab}). As it was reported in various news reports, the announcement was negatively perceived by many open source developers, who publicly encouraged other developers to migrate to \texttt{Gitlab} \cite{reutersgitlab}. Obviously this affected the community of bot programmers as well.

\begin{figure}[h!]
\centering
\includegraphics[width = 0.7\textwidth]{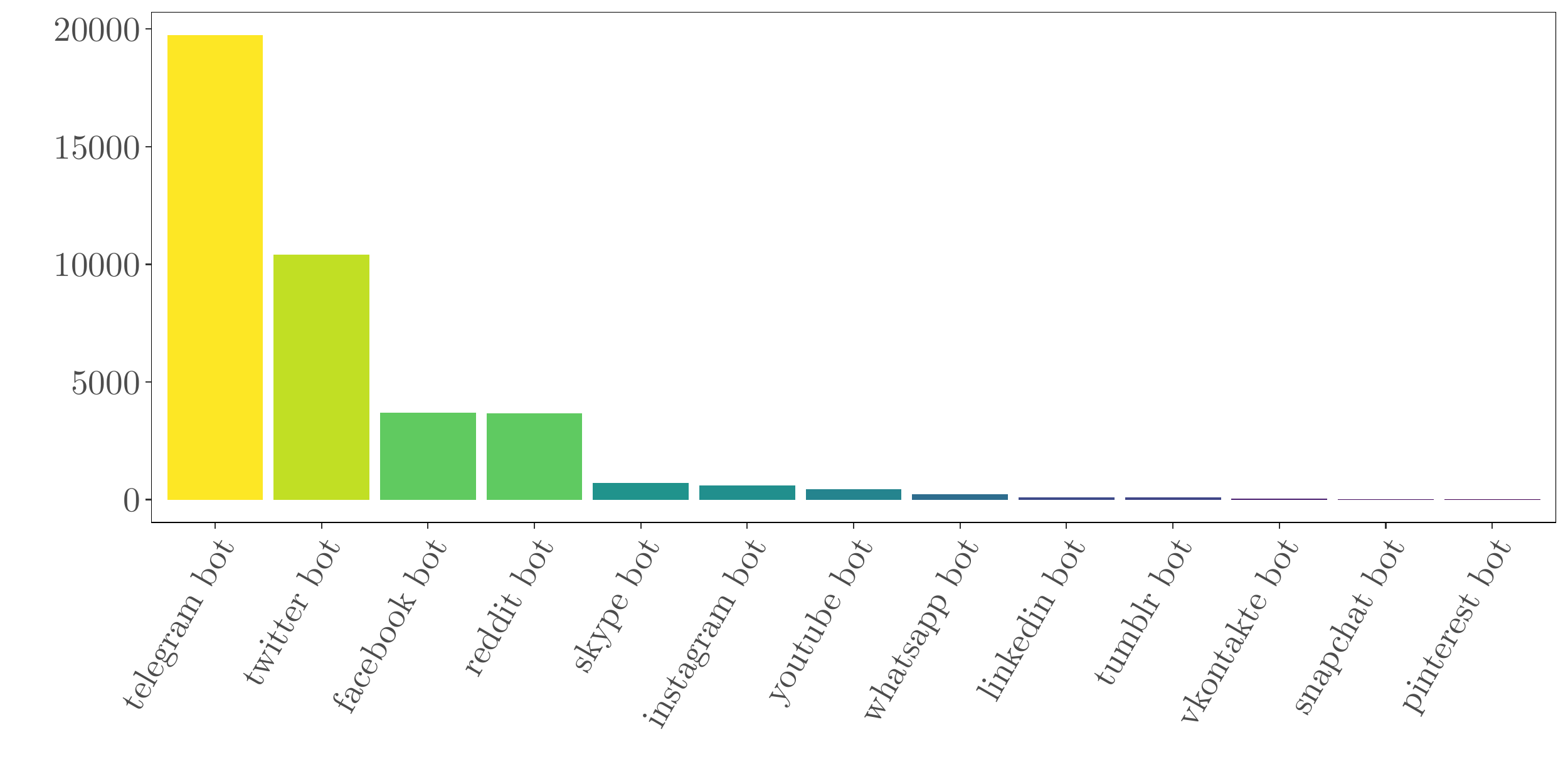}
\caption{Searchterm distribution for all social-media platforms}
\label{fig:searchterm_dist}
\end{figure}

\begin{figure}
\centering
\includegraphics[width=\textwidth]{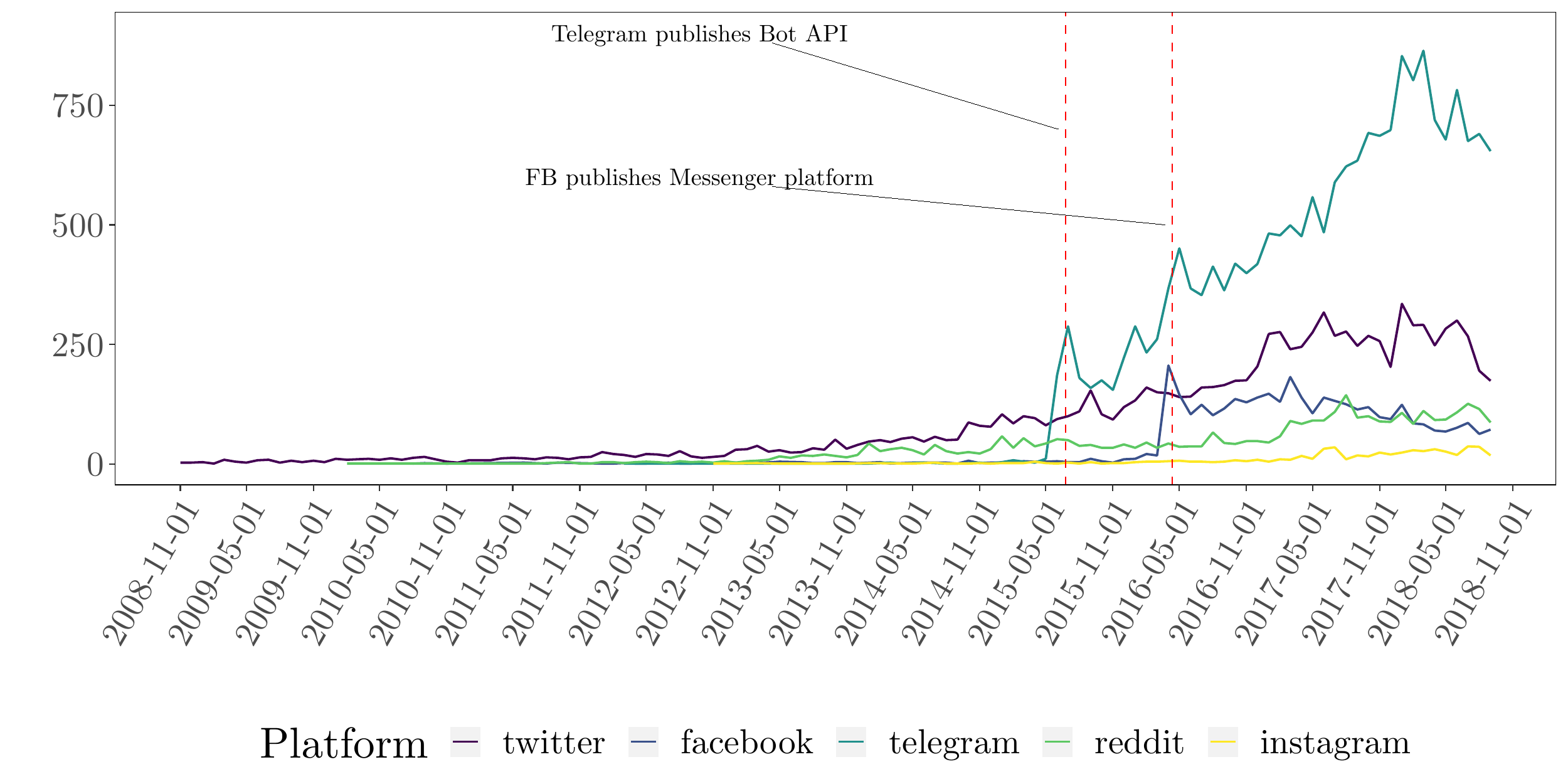}
\caption{Number of Repositories for different platforms over time}
\label{fig:telegramA}
\end{figure}

Over all collaboration platforms, we observed a similar distribution regarding the number of repositories for a specific social-media platform (Figure \ref{fig:searchterm_dist}). Most of the identified programs were produced for Telegram, followed by Twitter, Facebook and Reddit. At first sight this is a surprising result since Telegram is not considered as one of the big social-media players and the platform only exists since 2013. A detailed inspection of the creation date for Telegram oriented repositories revealed that until 2015 the platform did not receive a lot of attention. This changed in June and July 2015, when a significant increase in the number of related projects can be observed. We can directly explain this sudden increase by the fact that on June 24, 2015, Telegram officially launched it's open bot platform, making it easy for programmers to create automated bot programs via an external API. Furthermore, the functionality of creating inline bots (bots that can be addressed in any chat) led to a second raise of newly created applications in January 2016. Hence the social platform itself seems to directly impact the community of social-bot code.  Figure \ref{fig:telegramA} shows, amongst others, the number of newly created Telegram repositories over time.

\begin{figure}[h!]
\centering
\includegraphics[width=\textwidth]{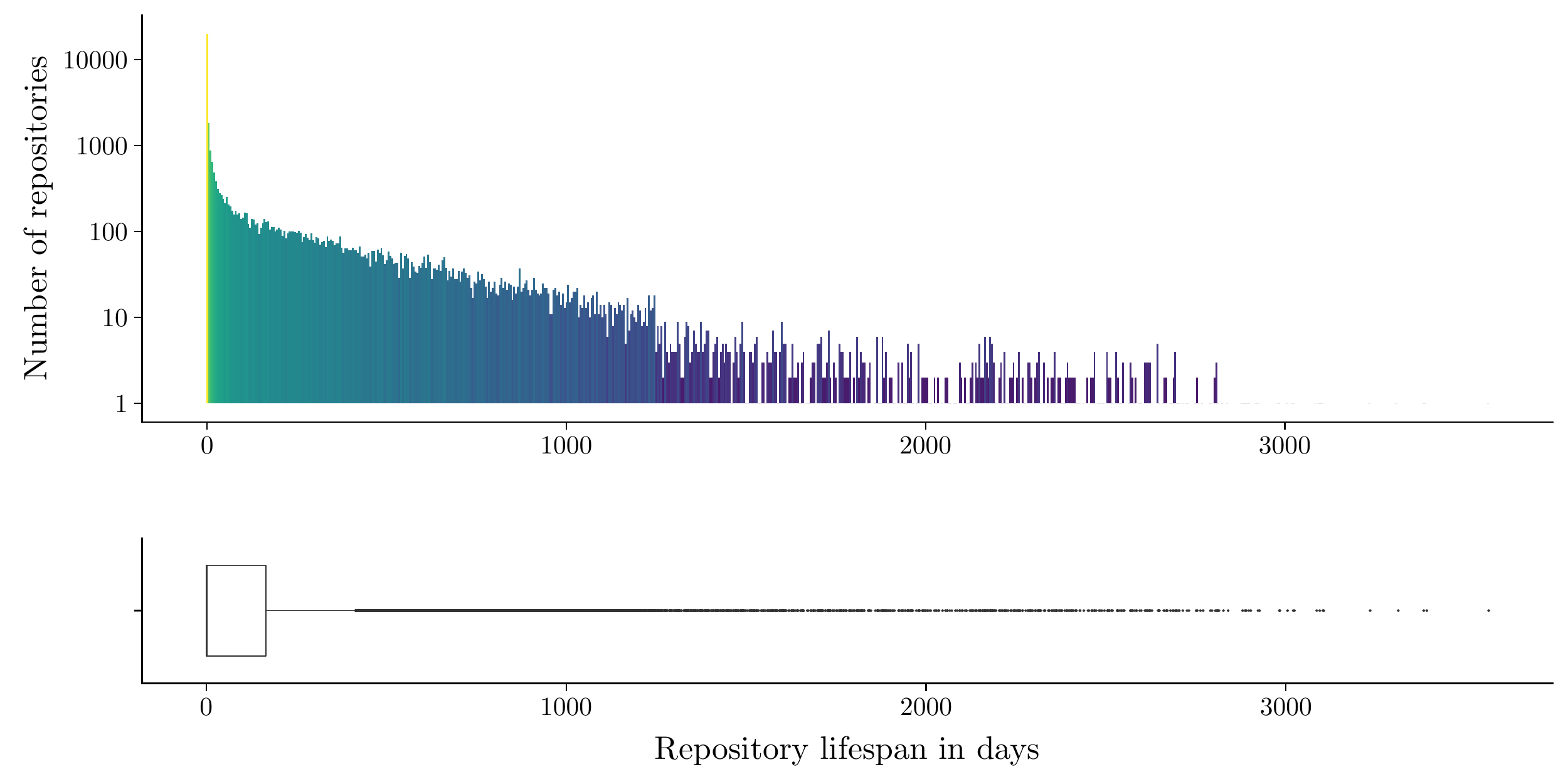}
\caption{Lifespan of the crawled repositories}
\label{fig:telegramB}
\end{figure}

In a second step we analyzed different lifespans of repositories. We define the lifespan of one single Repository as the time between the creation date and the last activity. Moreover an activity is characterized by any Repository interaction such as a new contribution, a fork or a newly assigned issue. We observe that more than 50\% of the crawled Repositories (18.000) have a lifespan of 0 days (Figure \ref{fig:telegramB}). This means that such Repositories were once created on a specific date and did not receive any further update after publication day. As indicated in \cite{Kollanyi2016}, some developers use the \texttt{Github} platform only as a medium for sharing their code rather than collaborating with other users. 

\paragraph{API Support and Programming languages}
%
%
Due to the heterogeneous structure and conflicting goals between different social media platforms, companies handle third party access to the service they provide in a different manner. Whilst some platforms actively encourage developers to create external applications by providing dedicated interfaces for accessing their data and functionality, some platforms do not offer such information. Within this work, we differentiate between four distinct classes of third party access. 
\begin{itemize}
\item Social media platforms that are assigned to the \texttt{BotAPI} class are considered the most favorable ones for programmers. Such platforms do not only offer API's for third party institutions, but also dedicated services and functionality for bot programs. Within our study Telegram was the only platform that provided such sophisticated interface. 
\item Platforms that offer an \texttt{API} to perform all the common tasks of the corresponding web-interface in an automated way. For most of the social platforms these tasks are following, creating comments and all kinds of flat social interactions (like/dislike). 
\item Platforms that offer a \texttt{limited API} access. Although an interface for third parties exists, the range of functionality is limited. Platforms that are assigned to this class, for example only allow to access private user data.

\item Platforms that offer \texttt{no API} and no other means of interface for external parties.
\end{itemize}

Most of the social media platforms of interest offered some kind of API to third parties. Only Whatsapp and Snapchat do not provide any official API interface. Not surprising, those platforms are situated on the lower ranks regarding the number of repositories found for the specific search term. Most of the platforms with a higher rank do offer a more sophisticated API. In general we observed a positive rank correlation between the number of repositories found for a specific social platform and the corresponding level of API support ($\rho=0.78$, see Table \ref{tab:support-correlation1}). Overall,  the \texttt{limited API} was the most prominent class to which social platforms are assigned. This can be explained by strict privacy policies of some bigger social-media platforms. Because of recent incidents, where private data was used for manipulation purposes (e.g. Cambridge Analytica), those companies were widely criticized by the public for providing unrestricted data-access. As a result, some platforms changed their policy and consequently limited their API functionality \cite{CamebridgeAPI}. Instagram, for example, only allows to access and analyze account related data and activities. In contrast to the past, external information like the access to the followers of arbitrary users is strictly prohibited, unless the target person explicitly gives the permission to acquire the desired information.

\begin{table}[tb]\small
\setlength{\tabcolsep}{9pt}
\begin{center}
\caption{API support for Social Platforms, sorted by number of repositories}
\label{tab:support-correlation1}
\begin{tabular}{@{}lcccc@{}}
\toprule
\textbf{Social Platform}& \texttt{no API}&\texttt{limited API} & \texttt{API} &  \texttt{BotAPI} \\ \midrule
Telegram           &             &  & & X   \\
Twitter            &              &  &X &   \\ 
Facebook           &              &  &X &   \\ 
Reddit            &             &  & X&   \\ 
Skype        &               &  &X &   \\ 
Instagram           &             &X  & &   \\ 
Youtube            &              &  X& &   \\ 
Whatsapp         &      X           &  & &   \\

Linkedin           &              &  & X &   \\ 
Tumblr        &              & X  & &   \\ 
vKontakte        &              & X &    \\
Snapchat        &  X            &  & &   \\
Pinterest            &             &X   & &   \\

\bottomrule

\end{tabular}
\end{center}

\end{table}

\begin{figure}[b]
\centering
\includegraphics[width=0.8\textwidth]{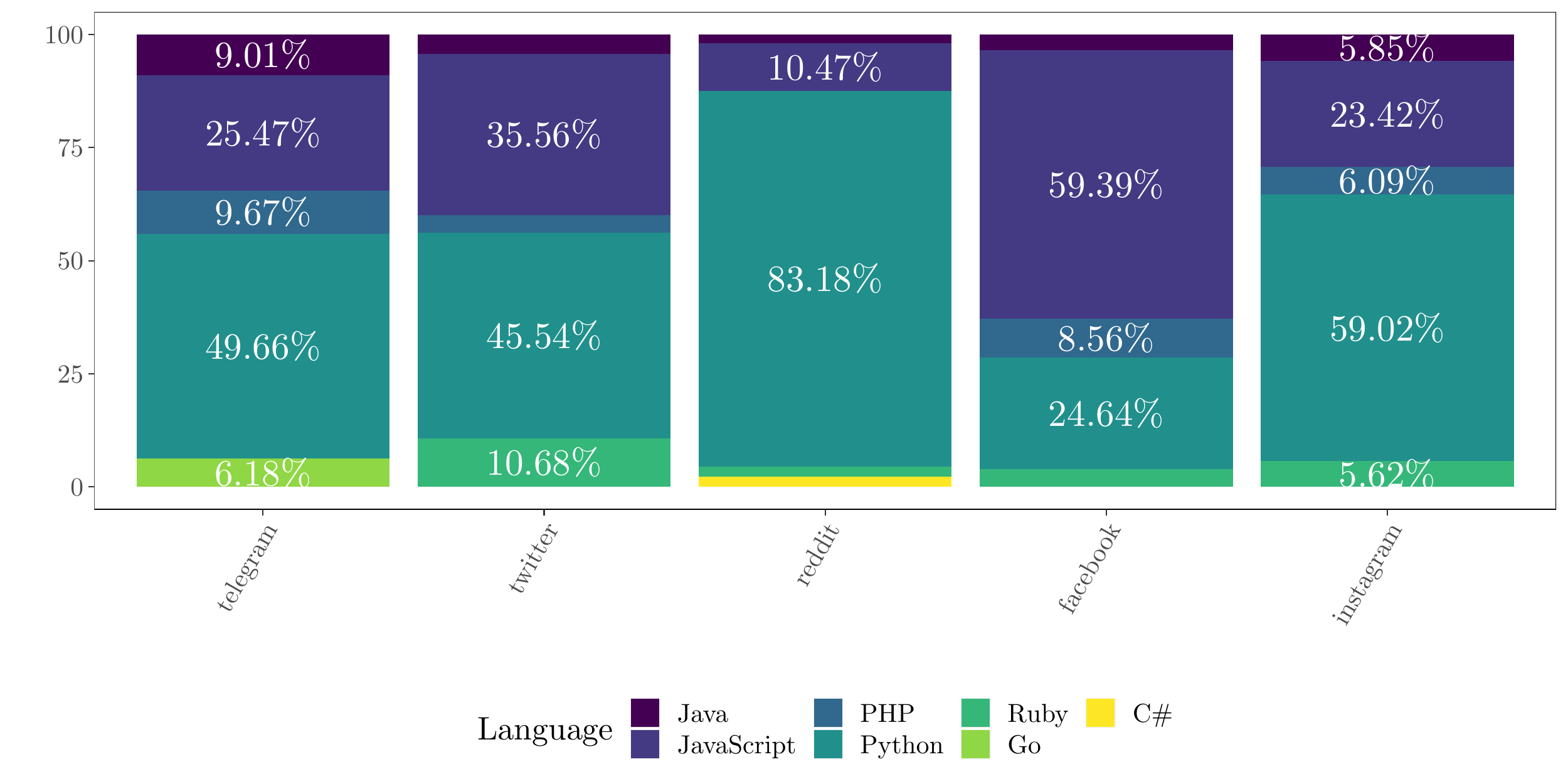}
\caption{Programming languages used for the Top 5 social media platforms}
\label{fig:languages}
\end{figure}

Most of the social media platforms which provide a dedicated API, offer some additional interfaces to access their service. These interfaces can be accessed by specific programming languages.  Within Figure \ref{fig:languages} we show an aggregated view of the repositories main programming languages for the top five social media platforms (in terms of repository count). The most common programming language over all platforms is Python. Interestingly JavaScript is also frequently utilized. While Facebook explicitly provides a Java Script Toolkit, this is not the case for the other platforms. In cases where the API is somehow restricted (e.g. companies privacy policy), programmers often directly access the web interface with JavaScript code to circumvent the official API.

\paragraph{Geospatial analysis:}
\texttt{Github} and \texttt{Gitlab} allow each user to specify their respective geolocation. As it was already indicated in \cite{Kollanyi2016}, the platforms use a free text field for collecting the information from the user. Therefore, we can not assume standardized data for our analysis. We utilized Google's geocoding API to receive longitude/latitude pairs that are approximately close to the location that was specified by the user. In total, we gathered information of 46.900 unique contributors, where geolocation information was present in 22.688 cases.

\begin{figure}[tb]
\centering
\includegraphics[width=\textwidth]{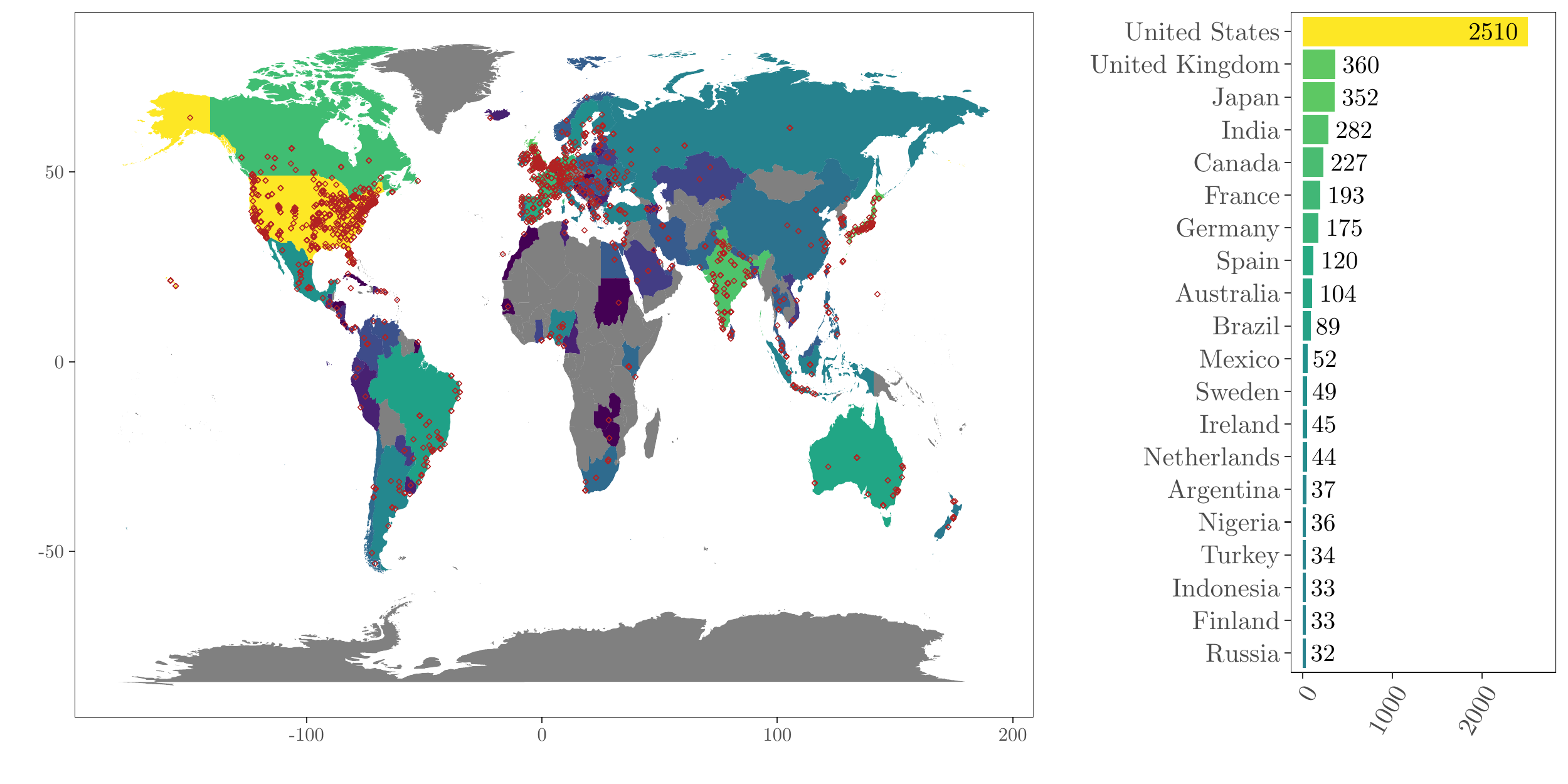}
\caption{Origin of Twitter repositories }
\label{fig:twitter}
\end{figure}

Kollanyi \cite{Kollanyi2016} already investigated the global distribution of twitter bot repositories. We present an updated version of the map with a similar distribution that was already observed in 2016. As one can see on the world map in Figure \ref{fig:twitter} the main part of the repositories belong to ten countries. This observation is comparable to the findings of \cite{Kollanyi2016}. Having a look at the top five most contributing countries and comparing them to the findings of \cite{Kollanyi2016} from 2016, shows, that the distribution basically stays the same. 

\begin{figure}[h!]
\centering
\includegraphics[width=\textwidth]{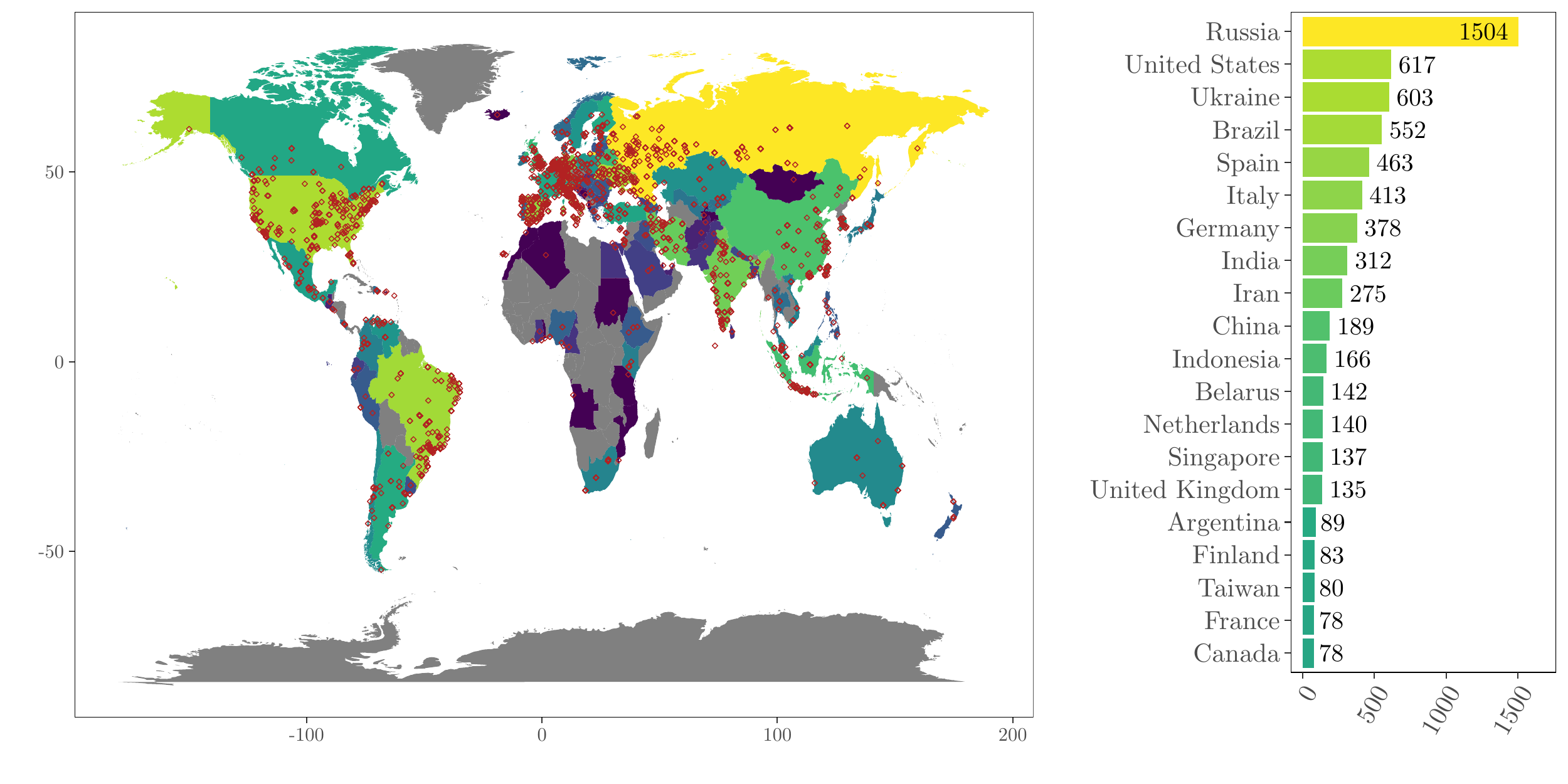}
\caption{Origin of Telegram repositories }
\label{fig:telegram}
\end{figure}

Most of the Twitter repositories originate from the United States. In \cite{Kollanyi2016}
the United States were directly followed by Japan.  Our updated version reveals that United Kingdom caught up to Japan and follows the U.S. by providing the second largest number of bot related repositories for Twitter. In contrast to Kollanyis study we also have access to location data of different social-media platforms. Directly compared to the distribution of Twitter, we observe some inherent dissimilarities between the platforms. While Russia does not play an important role in the context of Twitter bots, most of the Telegram bot code contributors are from that country (Figure \ref{fig:telegram}). A reason for this could be the popularity of Telegram within the Russian population~\cite{nytimestelegram}. 
Whatsapp contributors mainly originate from India, where the messenger is not only used for private communication, but serves as a central communication and information channel~\cite{StatistaIndia2016}.

\section{Content Analysis}
\label{sec:qual_analysis}

Identifying the capabilities, operational scenarios, and associated costs of automated bot programs in the context of social media platforms is regarded as one of the major goals of this study. Therefore -- on a qualitative level -- the content as well as the overall topics of available social bot code is of central interest.
Since, in general, manual inspection of the code base or the description of each repository is unfeasible for the number of gathered repositories, we utilize unsupervised learning techniques to identify topic trends over time in an exploratory way. We start with an additional and explicit content analysis of gathered repositories. We apply dominance filtering from decision theory in order to bound the number of 'interesting' repositories for manually detailed investigation.

\paragraph{Analysis of (interesting) Content:}
\label{sec:multiobjective}
To identify a subset of interesting repositories $R'\subseteq R$ for manual inspection, we specify three different indicators of interest: $N_c(r)$, the \emph{number of commits}, $L(r)$ the \emph{repository lifespan}, and $T(r)$ \emph{repository timeliness} with $r\in R$. While the number of commits is a rough indicator for overall user engagement in a software project (represented by the repository), the lifespan as defined in section~\ref{sec:descr_analysis} provides insights into the age of the respective repository. Further, we measure the timeliness of a repository as the difference between the current time and the time of the last update. Hence, a low timeliness value indicates a repository that has recently been updated, whereas a high value represents a repository that has not been updated for a long time and is probably inactive. 

Using the defined simple indicators, any repository $r\in R$ can be represented by a three dimensional vector $v(r) = (N_c(r), L(r), T(r))^T\in \mathbb{N}^3$. 

In our interpretation of these three indicators, an ideal (i.e. most interesting) repository for manual analysis would have large $N_c$, large $L$, and small $A$ components. Clearly, multiple dimensions of interest can result in incomparable repositories, such that no strict order of interest can be achieved. To identify the most promising trade-offs regarding the three indicators, we employ the notion of dominance, usually applied in the field of decision theory and multiobjective optimization. We say a $d$-dimensional vector $v$ dominates another $d$-dimensional vector $w$ (notation $v \prec w$), if and only if the following conditions hold: (1) $\forall i \in {1,\dots,d}$ : $v_i \leq w_i$ and (2) $\exists i \in {1,\dots,d}$ : $v_i < w_i$.
Note, that $\le$ and $<$ are interpreted as \emph{scalar domination}, i.e., that this relation is interpreted differently for maximization and minimization.\footnote{To overcome the distinction of minimization and maximization, we can convert minimizing indicators to maximizing indicators (or vice versa) due to the duality principle.}
In other words, a repository only dominates another repository, if for each indicator of interest, the first repository performs at least equally well as the other one and for at least one indicator strictly better. If this is not the case, we call both repositories incomparable or non-dominating to each other. 

To eventually select the dominant (and thus most interesting) repositories for manual content analysis, we perform a pairwise comparison of the indicator vectors of all repositories from $R$ and remove those, which are dominated (non-domination filtering). As a result, we acquire $83$ incomparable and non-dominated repositories forming $R'$ for further analysis.

The analysis is done manually, by reading through repository descriptions and read me files. The results show, that most of the repositories yield code for the social media platforms Telegram (33), Twitter (27), and Reddit (11), reflecting the overall distribution of all bot codes (see Figure \ref{fig:searchterm_dist}). There exist three repositories containing code for the platform Instagram and one repository per each of Whatsapp, Facebook, Youtube, and Skype\footnote{Five of the repositories were omitted, since they contain e.g. Chinese descriptions or are related to other topics.}.

Table \ref{tab:support-correlation} shows the functionality and purpose, which is handled by the repositories. Most of the repositories can be grouped, since they provide similar functionality. 

\begin{table}[!h]\small
\begin{center}
\begin{tabular}{@{}llr@{}}
\toprule
\textbf{Type}& \texttt{Functions}&\texttt{Number of Repos}  \\ \midrule
Native Social Bot Functionality           & posting, liking, &22\\ & following   &   \\
Wrapper/ Framework/ Library           & easy API access, &27\\ & pre-defined functions  &    \\
Other Service Functionality           & downloading data,& 14\\& payment handling, & \\ & include external services  &     \\
\bottomrule
\end{tabular}
\end{center}
\caption{Functionality of Top Repositories}
\label{tab:support-correlation}
\end{table}

Twenty-two of the top repositories contain code which enables the user to implement social bot activities. Exemplary actions are posting predefined content, building follower networks, or liking specific hashtags. Most of the repositories contain code, which enables users to run their own bot script by adjusting a few lines of code. Related to the bot taxonomy of Grimme et. al \cite{Grimme2017}, we  especially find bot repositories, which can be assigned to the class of simple bots. The 22 repositories, which are part of the two non-dominated layers, contain simple API supported functions like posting, liking or following. None of the code repositories provided code for "intelligently" acting bots. 

Apart from the native social bot functionality, the main share of repositories contains code in order to handle API access for a specific programming language or to provide predefined code libraries. These 27 repositories can be applied by users, which are willing to create their own bot programs. In contrast to the   repository type described above, these repositories do not provide off-the-shelf executable code for a specific bot program. Rather, they simplify the usage of programming interfaces for different programming languages. 

Next to frameworks and API wrappers, there exists a number of dedicated bot scripts. The scripts contain code, which enables the user to establish own bots on a specific platforms. The analyzed repositories provided bots, which are able to fulfill simple tasks like following, liking or sharing posts. Most often a detailed setup description is provided and the code is easy to customize for own purposes.

About 14 repositories provide code, which fulfills no common social bot functions. These repositories contain scripts, which expand the platform by additional functionality. Examples are functions for downloading data, adding payment options or building interfaces for the inclusion of external services like Open Street Map.

\paragraph{Topic Modeling:}
Topic-modeling is used in the context of natural language processing to automatically extract semantic meaning from a given data corpus by identifying similarities between documents and grouping them together. We consider the application of one of the most frequently used algorithm in the field of natural language processing: Latent Dirichlet Allocation (LDA)\cite{lda}. In LDA, a generative statistical model is used to describe documents as a set of latent topics, where each topic follows a unique word distribution, which in turn is generated by maximizing the conditional probability for a word to occur in a given topic. While LDA usually achieves excellent results for static corpora, i.e. data that does not change over time, it offers very limited insights when it comes to topic evolution over time. However, for the given analysis it is crucial to understand trends in the context of social bot creation and how and when these trends evolve.

Given the requirements mentioned before, we decide to additionally employ a different method for topic modelling which is able to track changes over time: Stream Clustering. While clustering in general is an unsupervised machine learning technique that aims to identify homogeneous groups of observations, stream clustering expands upon this idea and solves different shortcomings of the traditional approaches. First, traditional clustering algorithms have to iterate over the data multiple times. This is not feasible for large and potentially unbounded data sources. Furthermore, stream clustering algorithms provide mechanisms to decay clusters over time and therefore account for changes within the underlying data distribution (concept drift). In the context of this work, we utilize stream clustering mainly for dealing with concept drift and identifying trends within the bot creation community. Specifically, we interpret the crawled repositories as a discontinuous data stream over time and use the method \texttt{textClust} to cluster the repositories descriptions. 

Since the main goal of our analysis is the identification of time-depended trends and topics based on the repository description, a few pre-processing steps need to be conducted. In a first step, we select all repositories that have English descriptions with more than one word. Additionally, we remove all occurrences of social-media platform names from the documents, since the goal of the analysis is the identification of bot code functionality rather than platforms. Next, we apply common data transformation steps such as tokenization, stop word removal and lemmatization. In addition to unigrams, we also create bi-grams to account for close relationships between words within one document, such as word combinations like \textit{markov chain}. Based on the pre-processed data, we execute a LDA analysis~\cite{Rehurek2010}. Manual inspection of the results shows that about 15 topics are well suited to reveal existing repository types. Allowing a larger number of clusters leads to artificial separation of topics, while a smaller number of allowed clusters leads to overarching topics that contain multiple, semantically rather different repository types. 

\begin{table}\small
\centering
\caption{Top eleven (out of fifteen) topic representatives provided by LDA}
\label{tab:lda}
\begin{tabular}{@{}ll@{}}
\toprule
Topic & Words                                                                               \\ \midrule
0      & \textbf{chat}, message, group, send, user, app, via, google, friend, bot                     \\
1      & \textbf{API}, using, written, python, \textbf{framework}, library, php, create, use, written\_python  \\
2      & python, script, first, learning, small, price, twitterbot, bot, reddit, tutorial    \\
3      & tweet, random, reply, test, \textbf{markov}, text, chain, generates, markov\_chain, given    \\
4      & platform, slack, implementation, ruby, language, answer, question, messenger \\
5      & manage, aws, telegrambot, play, game, url, notification, lambda, world, live        \\
6      & \textbf{simple}, bot, weather, people, creating, thing, sample, heroku, program, template    \\
7      & tweet, word, every, sends, day, info, picture, give, hour, random                   \\
8      & news, game, user, service, follow, follower, help, automated, card, profile         \\
9      & post, nodejs, tweet, account, twitter, search, using, image, user, made             \\
10     & \textbf{chatbot}, example, quote, personal, based, schedule, assistant, daily, bot, working  \\\bottomrule
\end{tabular}
\end{table}

Table \ref{tab:lda} lists the resulting first eleven topics, which confirm our findings from the decision making and analysis process in Section \ref{sec:multiobjective}.  Most of the topics represent repositories that provide simple functionality or user action such as posting random content (e.g. images or predefined messages), linking videos, or following other users (5, 6, 7, 8, 9). Besides, some clusters describe more sophisticated functionality, which enables interaction between different accounts like chatting with other users (0, 3, 10). In this context, we observe that Markov chains play an important role. However, the top representatives of the resulting topics do not indicate that state of the art machine learning algorithms are utilized. Cluster one represents repositories that provide frameworks or use existing platform APIs.

\begin{figure}
\centering
\includegraphics[width=\textwidth]{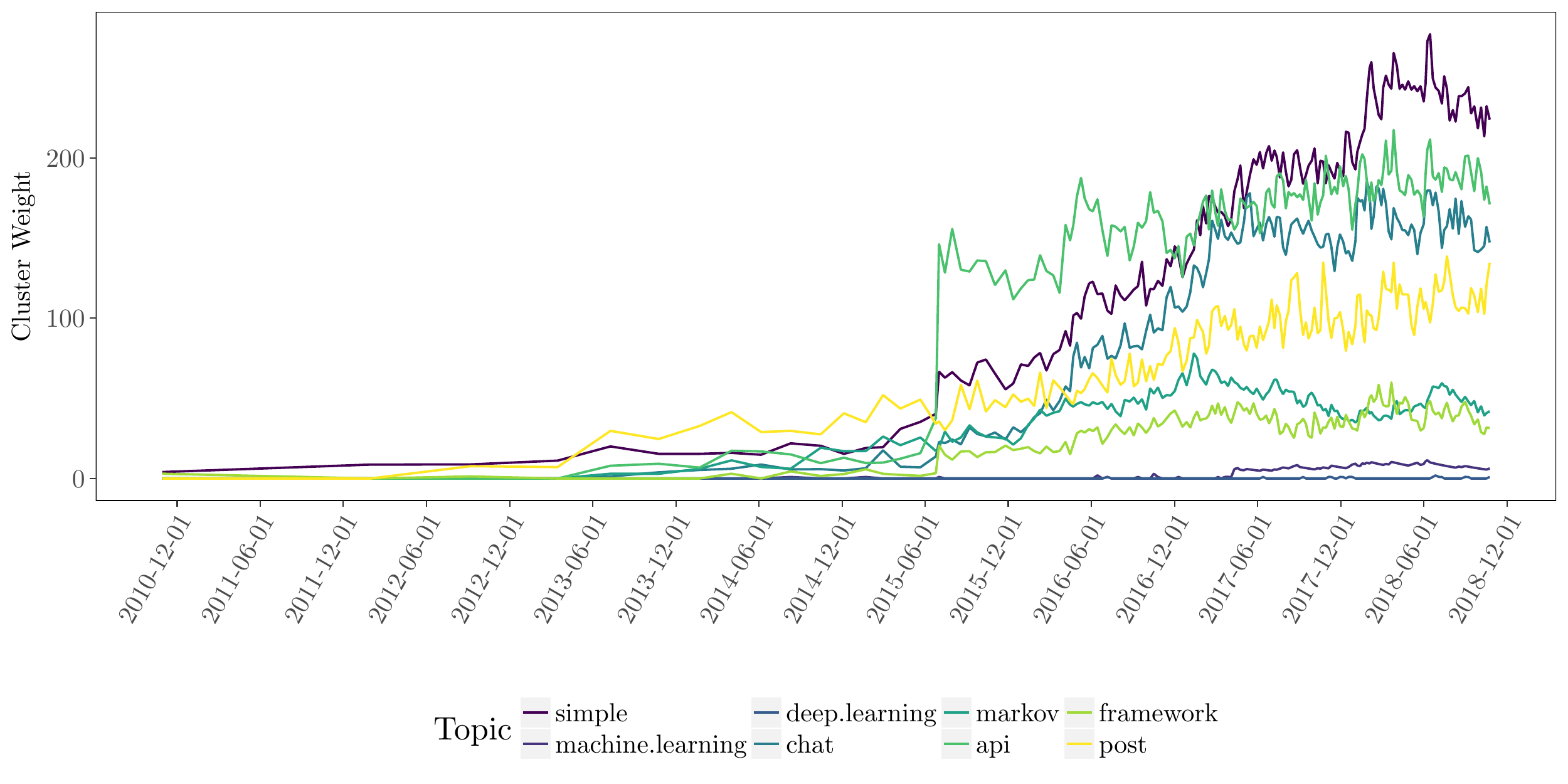}
\caption{Most important micro-cluster over time.}
\label{fig:micro-cluster}
\end{figure}


In a last step, we apply the stream clustering approach to inspect how the observed repository types behave over time, i.e. reveal potential trends within the bot-code community. As already mentioned before, we utilize a dedicated text-based stream clustering algorithm called \texttt{textClust}\cite{ER17Twitch}.
In general, the algorithm follows a two step approach, which is commonly used in the context of stream clustering. First, an online component summarizes the incoming text stream (represented as numerical \texttt{TF-IDF} vectors) in an online fashion, which results in a set of micro-clusters. These micro-clusters are considered as dense areas in data space. Periodically, micro-clusters are revisited and older entries which are not updated recently are decayed and ultimately removed. While micro-clusters are created, updated, and removed during the online-phase, they are aggregated macro level during the re-clustering phase. Therefore, traditional clustering algorithms can be used to aggregate the different micro-clusters. For our analysis, however, we focus on micro-clusters since they represent topics of repositories in our case. The algorithm is executed with default parameter settings. A pseudo-code description of the update and cleanup procedures of \texttt{TextClust} is provided in the Apendix, see Algorithms~\ref{Algorithm:update} and~\ref{Algorithm:cleanup}. Since the data-input is not uniformly distributed over time, we also select the algorithm's time variant, which fades data based on a predefined timedelta, instead of the number of observations. An in-depth description of the algorithm can be inspected in \cite{ER17Twitch}\footnote{The code can be accessed at \url{https://wiwi-gitlab.uni-muenster.de/stream/textClust}}. 

\begin{figure}
\centering
\includegraphics[width=0.6\textwidth]{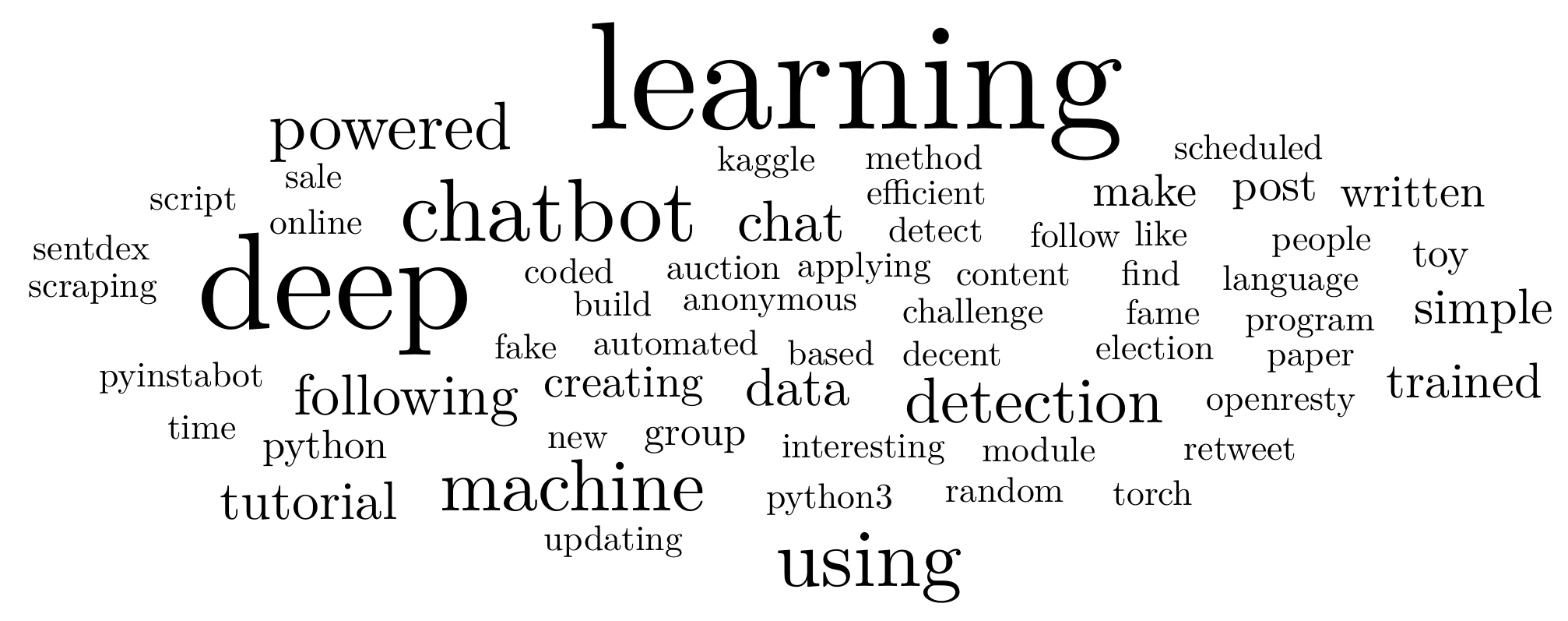}
\caption{Most frequent tokens in micro-clusters containing the term \texttt{deep learning}}
\label{fig:deep1}
\end{figure}

Within Figure \ref{fig:micro-cluster} the weight of micro-clusters containing specific terms  are displayed. In addition to important topic terms from the LDA analysis, we also inspected micro-clusters containing machine learning related tokens. The results of our stream clustering approach give some indication of topic importance (in terms of cluster weight over time). Micro-clusters containing the \textit{simple} term have the highest cumulative weight and therefore are dominant topics during the observed time-span. As it was already discussed, repositories that are assigned to those micro-clusters mainly implement simple functionalities such as automatic following, posting or displaying the weather forecast using external data. Clearly, machine learning repositories only play a minor role with a low overall weight, compared to the other cluster terms and represent a rather new trend that arose in early 2017. While most micro-clusters increase their weight over time in a rather linear fashion, the API term makes an exception with a significant weight increase in July 2015. As it was already analyzed before, this is due to the fact that Telegram published their bot API, one month before. Hence, we see that external events can directly impact the bot community. Since machine/deep learning was not reflected as a distinct topic within our LDA analysis (due to the small number of repositories containing these terms), we use our  stream clustering results to get an understanding of important terms, related to that topic. Therefore we look at micro-clusters containing either the term machine learning or deep learning and all tokens that are assigned to the similar cluster. Figure \ref{fig:deep1} shows the results as a word cloud. It indicates that machine learning techniques are mainly used in context of chat bots, i.e. the creation of human-like messages. However, also detection mechanisms such as hate-speech identification are present. 

Furthermore, we manually extracted and analyzed repositories, which contain keywords like machine learning, deep learning or artificial intelligence within their description. As a result we identified and inspected 85 distinct repositories manually. However, having a look at the  descriptions and read-me files, it turns out, that most of the machine-learning-related repositories (17) contain code for social bot detection. The contributors report on using classification approaches like Support Vector Machines or Logistic Regression in order to distinguish between human and bot steered accounts. Next to bot detection, content detection, e.g., in terms of detecting hateful or harmful content, is covered by the analyzed repositories (14). 

A large number of repositories (15) is related to chat functionality. Contributors use machine learning, deep learning and AI techniques to create chat bots, which are able to automatically answer to messages. Other repositories (14) create new posts on basis of former posts or an external text data base. Here natural language processing techniques are used to imitate specific writing styles.

Only two of the 85 repositories are described as "stand alone" bot scripts, which could be implemented to replace a human user. It is important to mention, that those scripts do not provide code for intelligent bots like introduced in e.g. the taxonomy of Grimme et al. Rather, they include some language processing techniques, which e.g. imitate a specific writing style for posts, or filter by the use of e.g. classification for interesting content to follow/like/share.

Next to the described functionality, there exist a number of repositories (8), which simply name keywords like "Machine Learning" or "AI" within their description, but do not contain machine learning, deep learning or AI technologies at all.
The remaining 15 repositories are either not further described, are already outdated, or contain descriptions of other languages. 

\section{Discussion}
\label{sec:discussion}
With this work, we provide a comprehensive study on the availability and state of development of open program code, which can be utilized for realizing the automation of accounts in social media. Based on developer repositories in the most relevant open collaboration platforms (mainly Github, Gitlab, and Bitbucket), we analyzed software development for a broad spectrum of social media platforms using various knowledge discovery techniques that range from descriptive analysis to stream clustering for trend detection. While the descriptive analysis proves overall availability of projects dealing with the development of social bots, a partition of data with respect to social media platforms provided additional insights into the importance and accessibility (and thus possible resilience) of these platforms for (against) simple automation of accounts. The performed geospatial analysis offers insights into regional importance of platforms and may point to main targets for social bot application at these locations. This results in different levels of potential "threat" in the world for different platforms as a consequence of user preferences. A specific view on programming languages, however, indirectly tells us some details on the operators and programmers of social bots. The overwhelmingly dominant usage of interpreted programming languages (Python and Java Script) allow simple and rapid development of code also for inexperienced developers. This suggests that a large proportion of the observed projects is developed to enable or simplify usage of social bots  or the realization of data acquisition from social media platforms at a prototype level. Additionally, the availability of simple API frameworks for specific social media platforms implies the use of simple programming languages like Python, while Java Script is presumably used to access data and functionality which is not reachable via the APIs.

The qualitative analysis of social bot projects is based on relevance of repositories, which was determined by a decision theoretical approach considering multi-criteria domination regarding repository life span, timeliness, and number of commits. The analysis of these repositories showed that the predominant functionalities provided in those repositories are restricted to simple bot development tools, frameworks, or specific services not specifically related to social bots alone. 

While these findings suggest, that openly available social bot technology is of rather simple nature, the topic modelling via LDA and stream clustering provides deeper insights into relevant types of social bot-related development as well as into the change of development focus over time. These aspects specifically address research question two of this work: We do not find ready-to-use software that exceeds simple functionality like (re-)posting prepared content, favoring articles/posts, or following other accounts in an automated manner. Also, the most important topics extracted by stream clustering suggest, that further functionality is restricted to building blocks for social bot construction, chat bot facilities, and API wrappers or alternatives / bypasses. Interestingly, machine learning (related to artificial intelligence) is an existing but neglectable topic in open social bot development. It mainly appears in the context of social bot detection. Sophisticated or ready-to-use mechanisms for the implementation of 'intelligent bots' are not available.

The observations of the descriptive and qualitative analysis together imply that we can distinguish two scenarios for the costs of social bot development (third research question). The simple scenario is the one where operators only aim for trivial usage of the available APIs of social media platforms performing actions like posting, favouring, or sharing. For these purposes, off-the-shelf software is freely available and feasible. Additionally, there is a set of proprietary interfaces and frameworks to easily enable such tasks. Considering very simple cases of content distribution, the realization of this type of social bots is relatively cheap. More expensive but probably in most cases still feasible is the realization of advanced social bots that simulate human behavior on the activity level - not in interaction with others. Available and open (some even well established and continuously maintained) social bot frameworks enable the enrichment of ready-to-use building blocks with more complex code. Experiments of Grimme et al.~\cite{Grimme2017,Grimme2018} demonstrated, that such extensions are feasible and require only medium technical expertise. A major gap, however, can be identified between the sometimes postulated existence of intelligently acting bots -- i.e. bots which are able to produce original content (related to a defined topic), provide reasonable answer to comments, or even discuss topics with other users -- and available open software components. The absence of tools implies that either the development of such techniques is too difficult and too costly to be provided open for the public, or that these techniques are scientifically too advanced to be subject of current social bot development. Current reports on experimental intelligent social bots by large software companies, however, suggest that there is currently no productive 'intelligent' software available~\cite{wptay2016,wiredzo2017}. Others observations show that 'intelligence' on conversational level is (intentionally) restricted to advanced chat bot capabilities~\cite{quartzzo2018}, as simple learning approaches are too sensitive regarding external manipulation. As such, the costs for realizing 'intelligent' bots can be considered infeasible, today.

Overall, this work and its analysis provides a new foundation for further discussion on the existence and influence of social bots extracted from an extensive evaluation of data on current open software projects. Certainly, software development is an ongoing process and technologies will change or advance together with advances in science (e.g. new developments in artificial intelligence) or with modifications in the technological environment (e.g. changes in APIs or accessibility of social media platforms). This paper provides a methodological framework for comprehensive analysis that goes beyond a singular investigation. It can be applied over time to monitor the development of technology and application of social bots.

\paragraph{Limitations and Future Work:}
Despite from the fact that our work reveals comprehensive insights into the bot programming community, some limitations still exist. One of the main limitations of this work is the restriction on open-bot code. It could be argued that social bots are frequently used in context of user manipulation. Therefore sophisticated bot programs which work well on that manipulation task, exhibit potential monetary value and thus could be sold to interested third parties. Hence, a platform which is based upon the idea of open-source code distribution may not be the right place to host the program code. In order to overcome this issue, a supplementary study, focusing on dedicated marketplaces in the dark web, could be conducted. Another limitation of our study is missing functionality verification. Since the source code of the different repositories was not explicitly executed, we cannot verify whether the code actually implements the functionality that was described in the repository's metadata. While for simple functionality such as automatic procedures to follow persons or liking their posts, we expect that the description actually reflects the described services, we cannot expect this from repositories that do apply complex machine learning algorithms and offer full-service bot programs. Lastly, we may not captured all social bot code that is currently available on the open-source platforms because they were not found by our predefined query. 

As we showed within our analysis, the community of bot code is constantly changing over time. Bot code programmers highly depend on the social platforms and their support for external programming interfaces while the platforms themselves constantly adjusting their policy and therefore also their API's because of public pressure. Furthermore new technologies such as machine- or deep learning emerge in the field and are more frequently utilized. Although we currently do not observe any sign that intelligent bots, as they were described in Section \ref{sec:related_work}, we do see a need for monitoring the development regularly. 
\newpage
\bibliographystyle{ieeetr}
\bibliography{literature}  

\section*{Appendix}
\begin{figure}[h!]
\centering
\includegraphics[width=\textwidth]{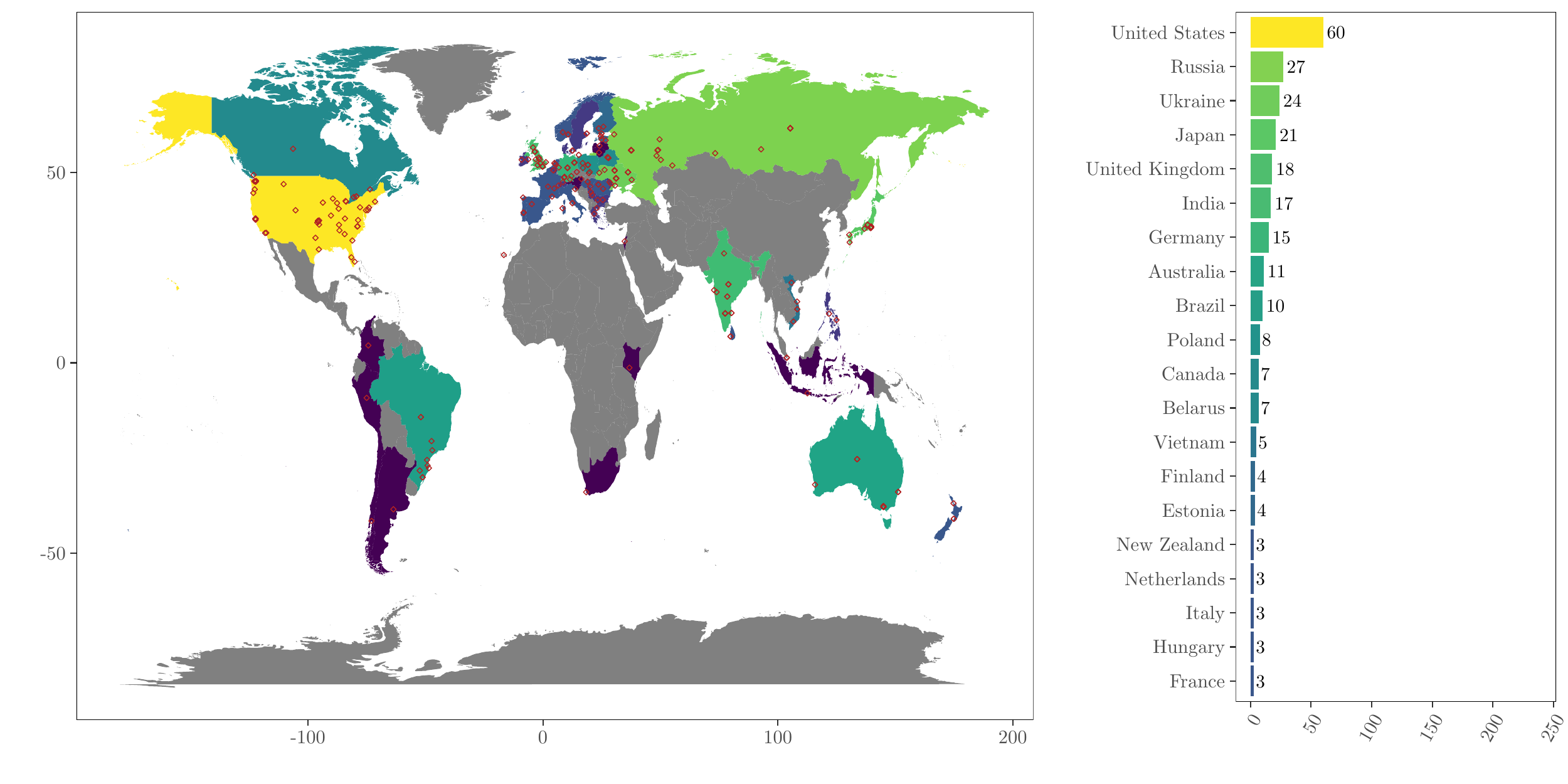}
\caption{Origin of Skype repositories.}
\label{fig:geo-skype}
\end{figure}

\begin{figure}[h!]
\centering
\includegraphics[width=\textwidth]{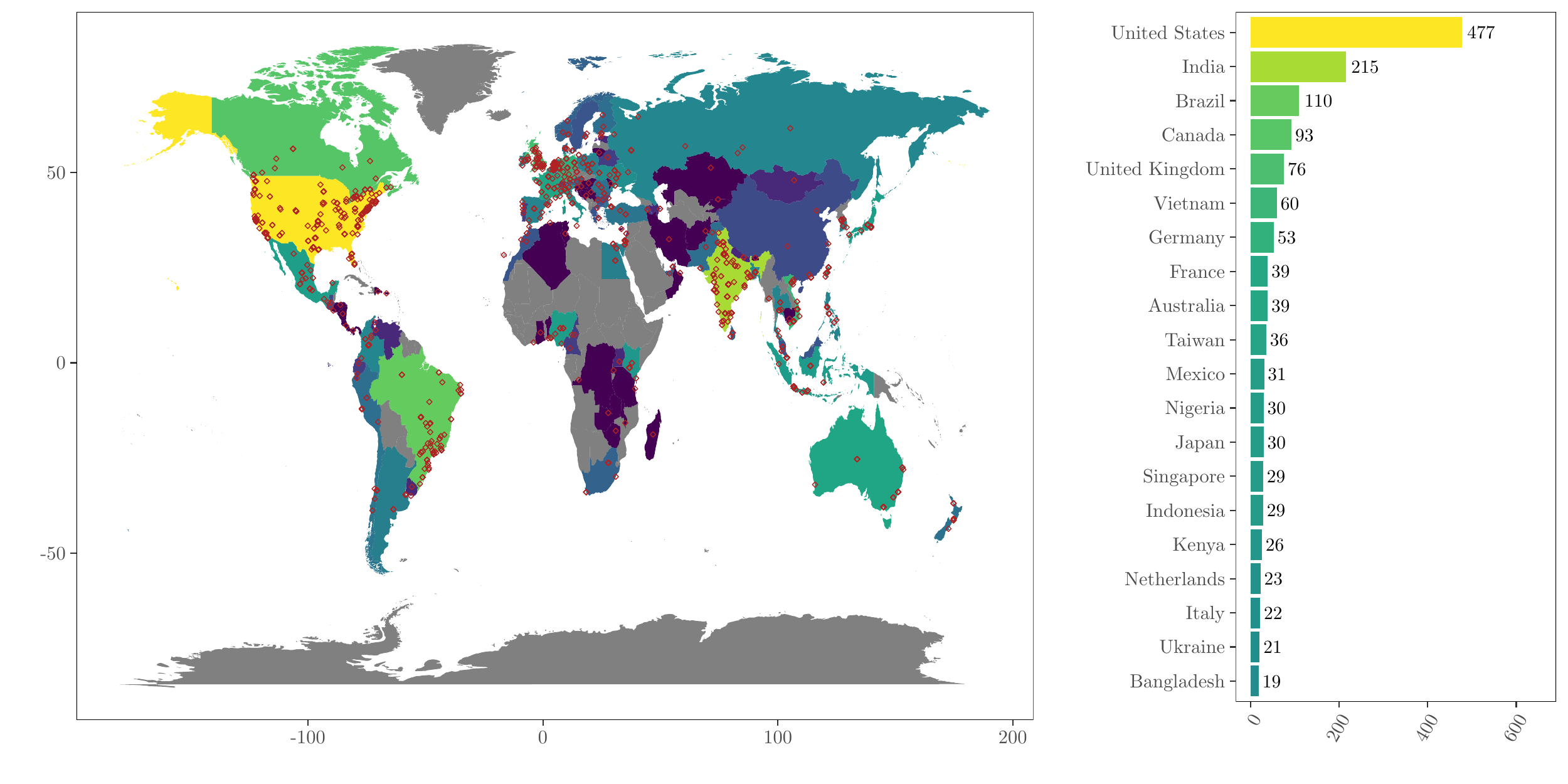}
\caption{Origin of Facebook repositories.}
\label{fig:geo-facebook}
\end{figure}

\begin{figure}[h!]
\centering
\includegraphics[width=\textwidth]{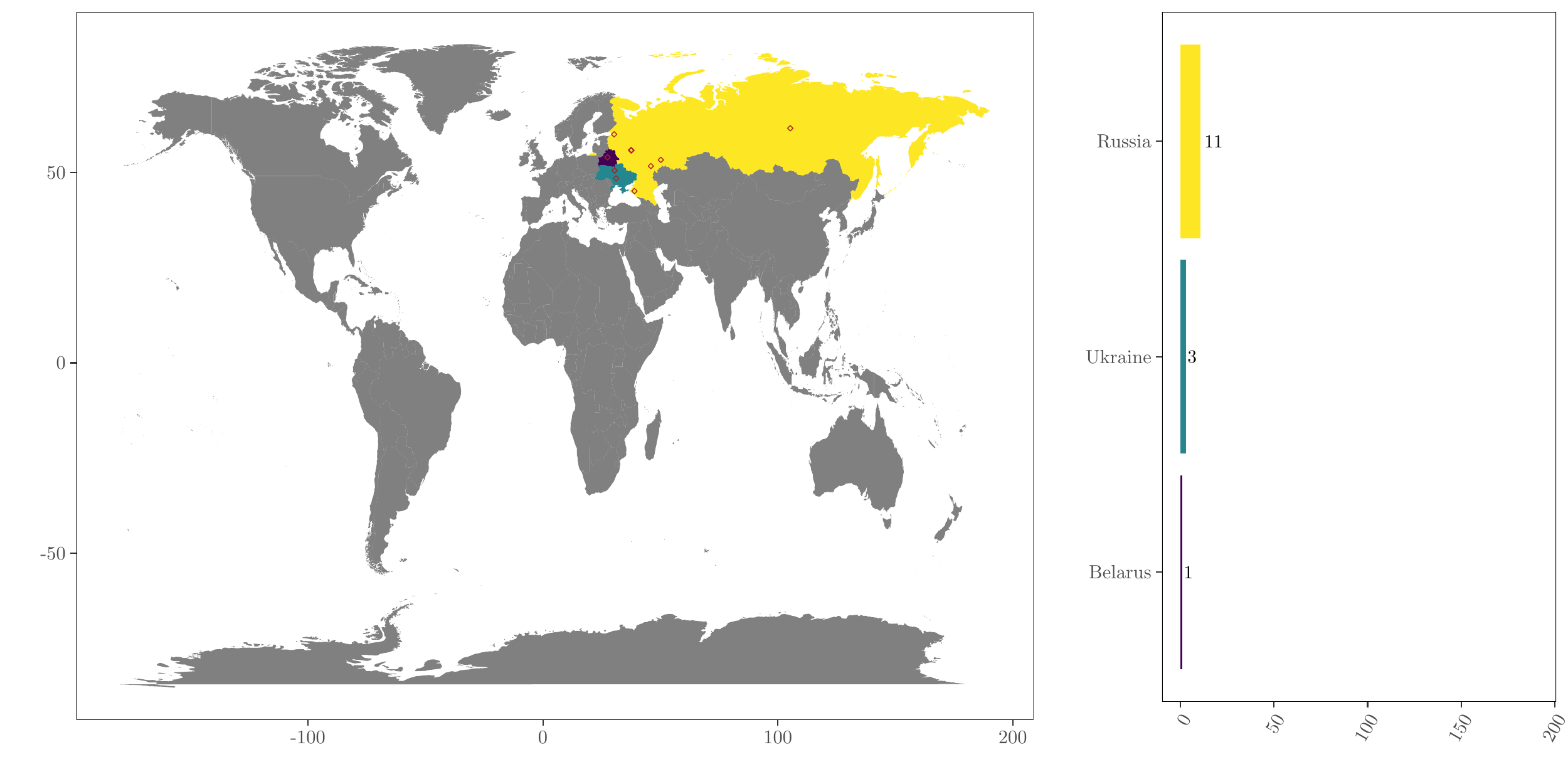}
\caption{Origin of vKontakte repositories.}
\label{fig:geo-vKontakte}
\end{figure}

\begin{figure}[h!]
\centering
\includegraphics[width=\textwidth]{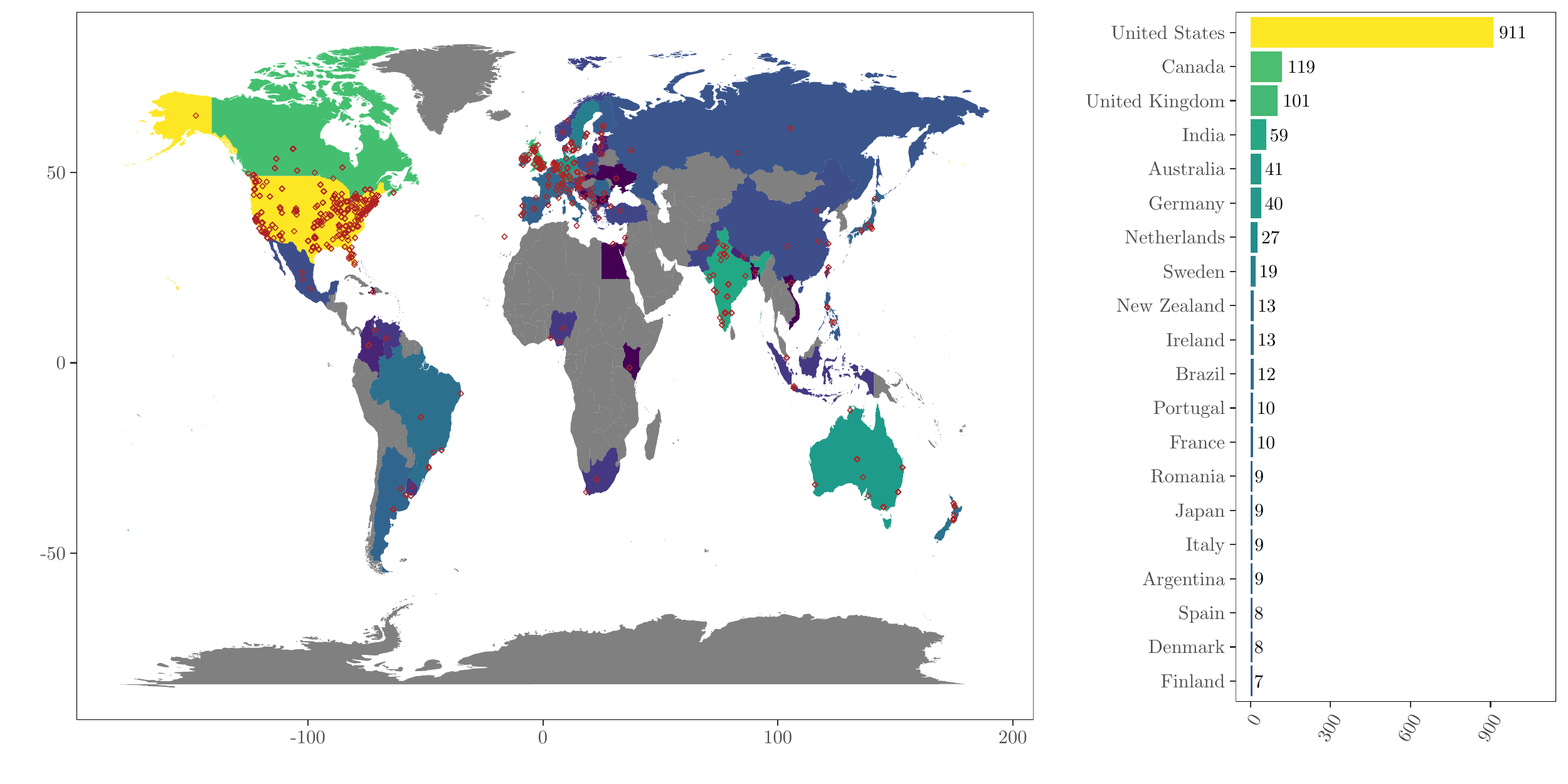}
\caption{Origin of Reddit repositories.}
\label{fig:geo-reddit}
\end{figure}

\begin{figure}[h!]
\centering
\includegraphics[width=\textwidth]{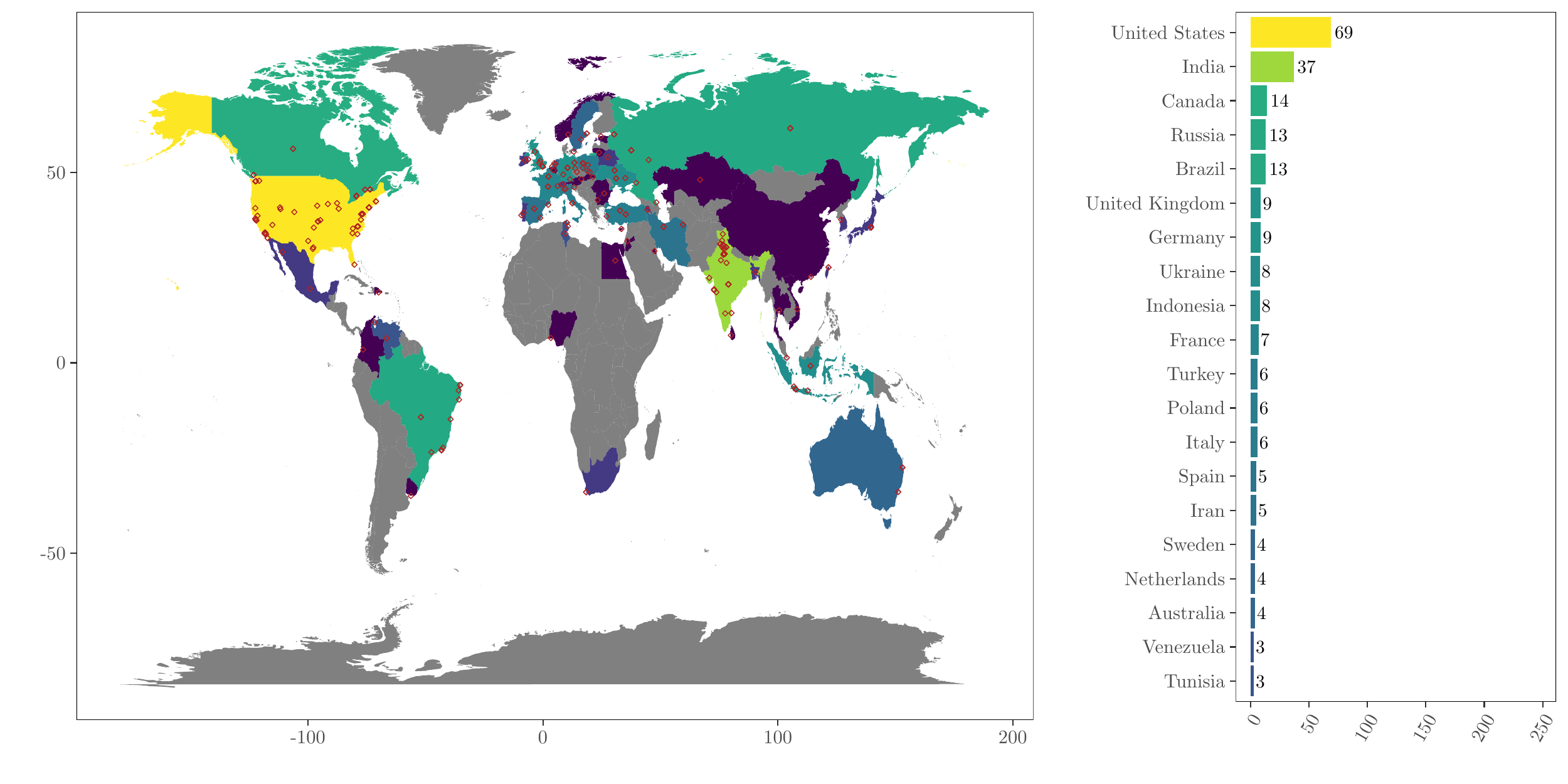}
\caption{Origin of Instagram repositories.}
\label{fig:geo-insta}
\end{figure}

\begin{figure}[h!]
\centering
\includegraphics[width=\textwidth]{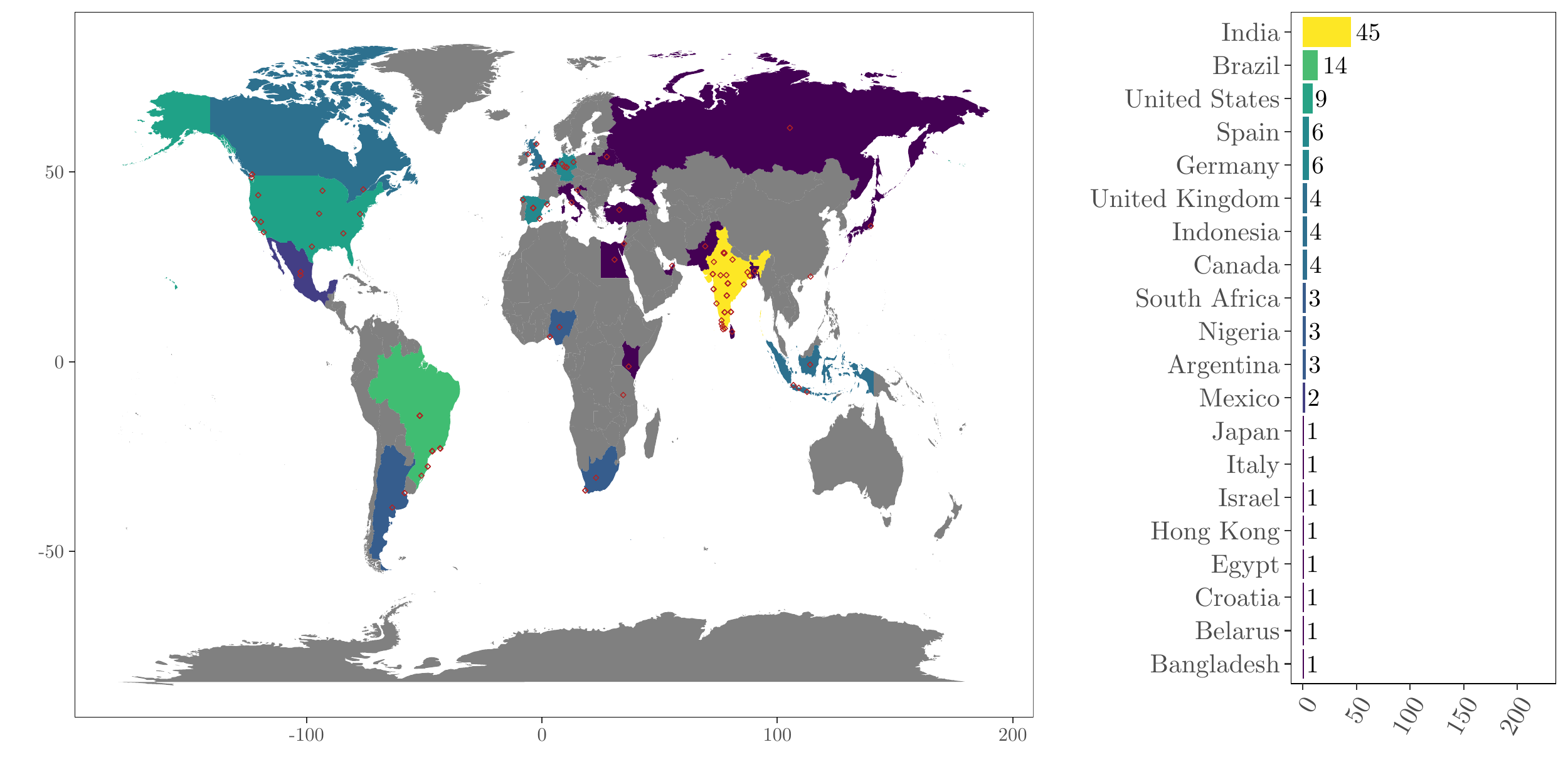}
\caption{Origin of Whatsapp repositories.}
\label{fig:geo-whatsapp}
\end{figure}

\begin{figure}[h!]
\centering
\includegraphics[width=\textwidth]{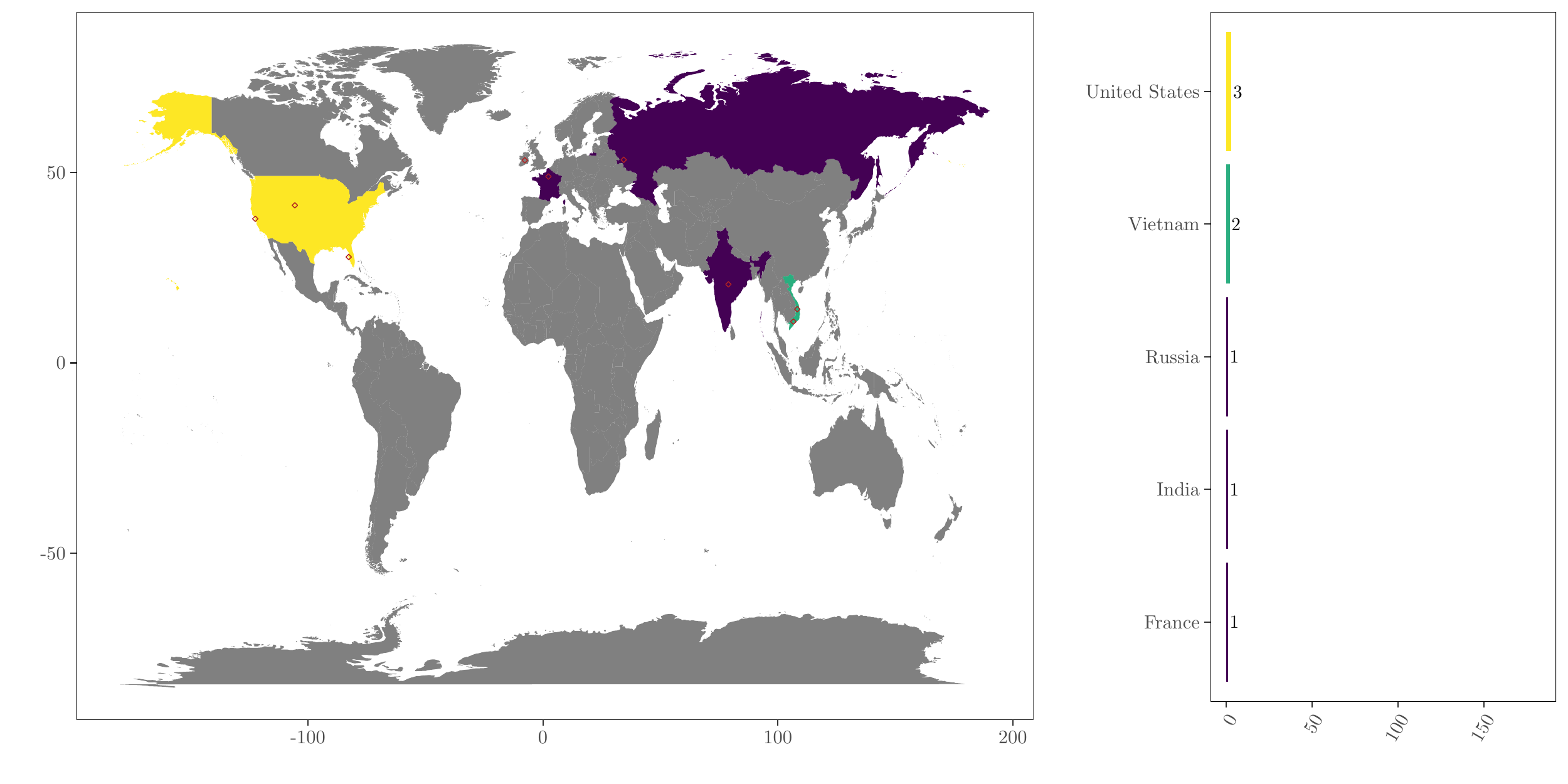}
\caption{Origin of Pinterest repositories.}
\label{fig:geo-pinterest}
\end{figure}

\begin{figure}[h!]
\centering
\includegraphics[width=\textwidth]{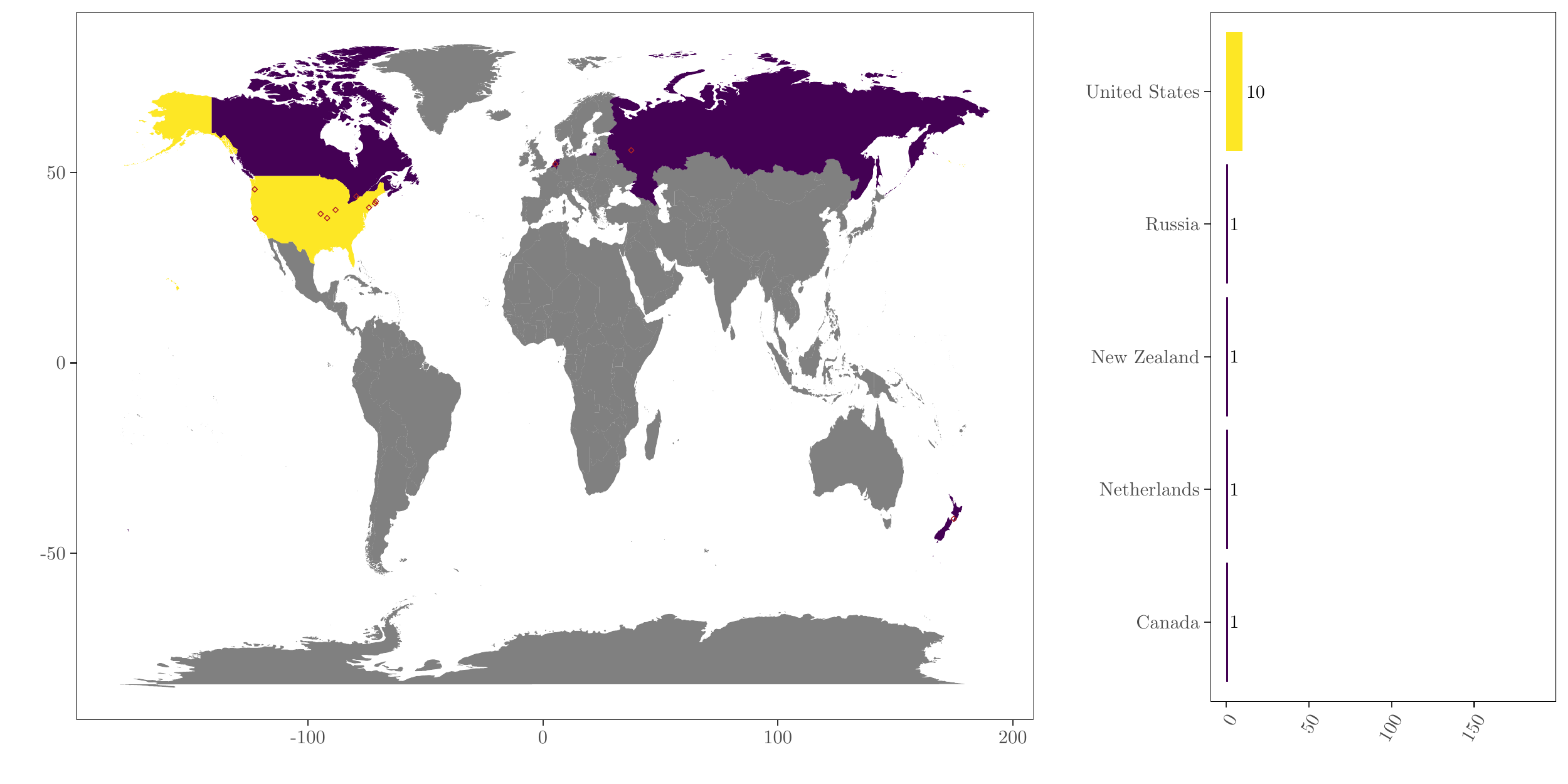}
\caption{Origin of Snapchat repositories.}
\label{fig:geo-snap}
\end{figure}

\begin{figure}[h!]
\centering
\includegraphics[width=\textwidth]{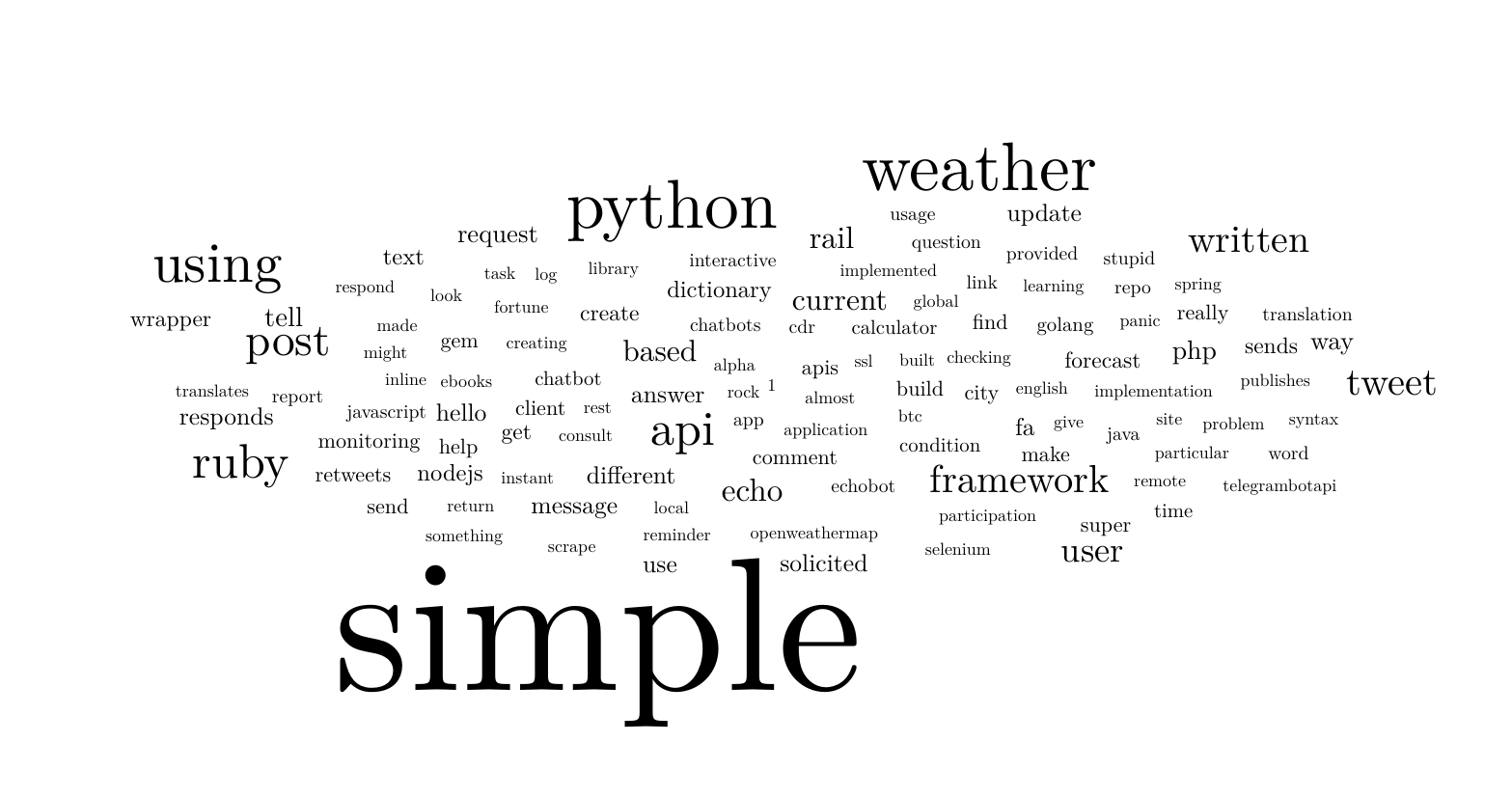}
\caption{Word cloud for repositories containing the search term "simple".}
\label{fig:wc-simple}
\end{figure}

\begin{figure}[h!]
\centering
\includegraphics[width=\textwidth]{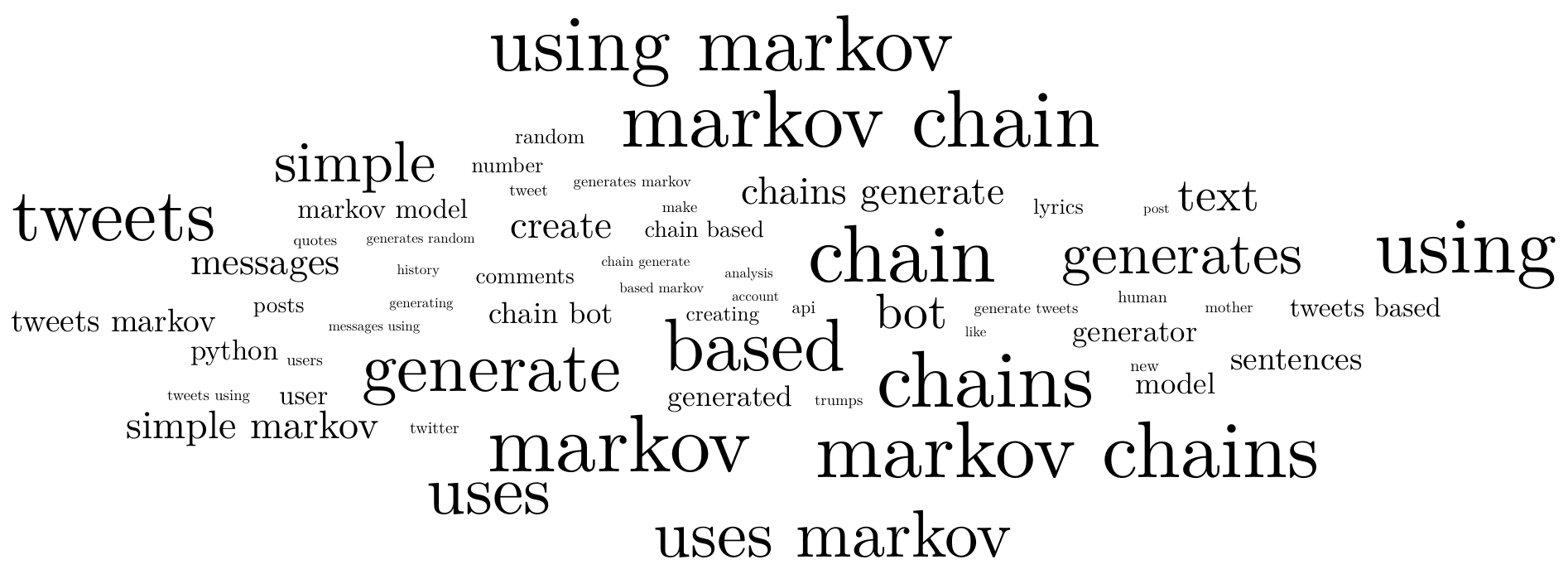}
\caption{Word cloud the repositories containing the search term "markov chain".}
\label{fig:wc-markov}
\end{figure}

This appendix contains a schema of the data acqusition process as well as supplementary material from the descriptive analysis and the topic modeling approach. The provided Figures~\ref{fig:geo-skype} to~\ref{fig:geo-snap} show the geospacial analysis of repositories, while Figures~\ref{fig:wc-simple} up to~\ref{fig:wc-markov} detail the determined topics from the application of TextClust as word clouds.

\begin{figure}[h!]
\centering
\includegraphics[width=\textwidth]{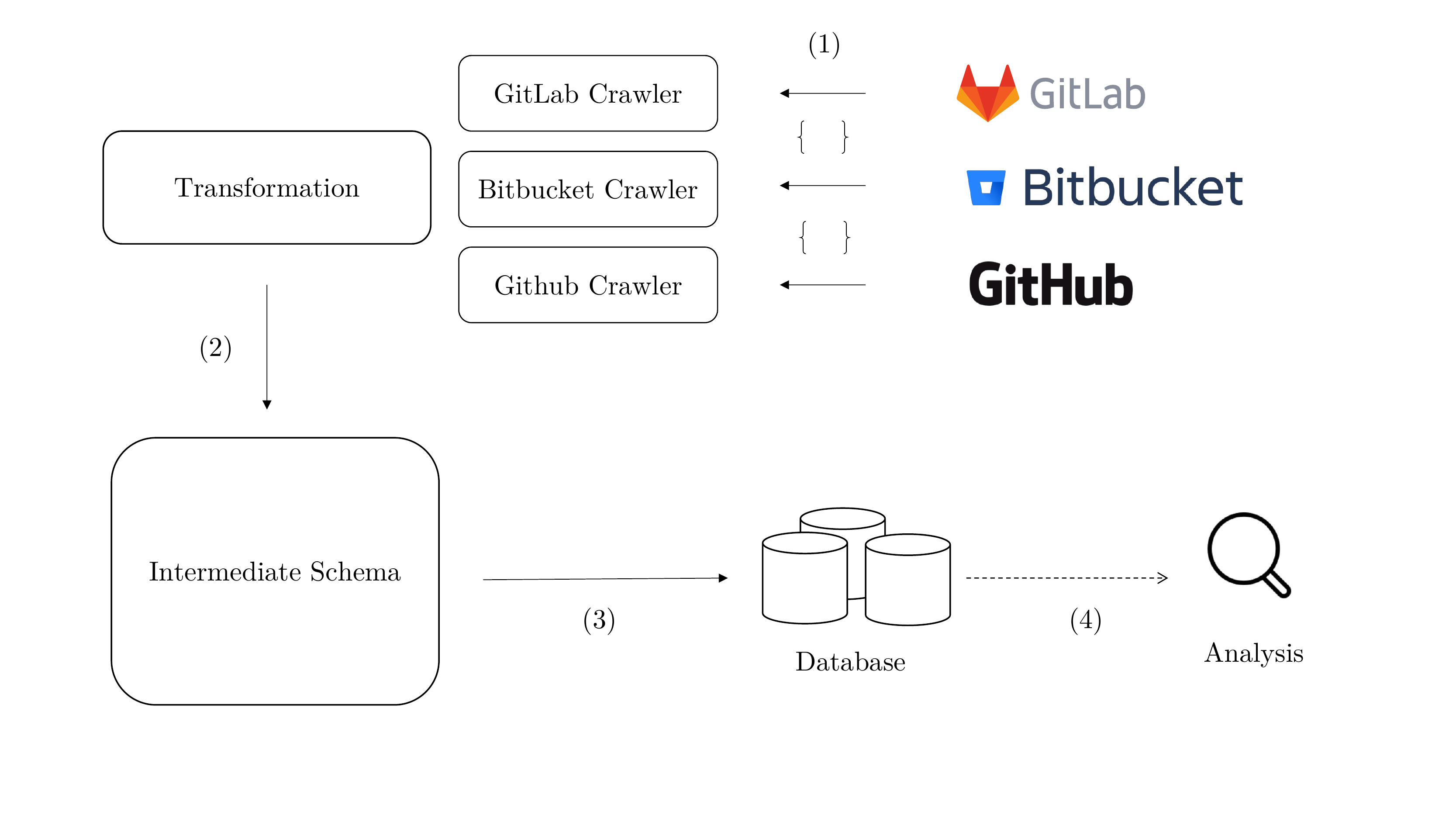}
\caption{Data Aquisition process and architecture with individual crawling instances (1), transformation of heterogeneous data into an intermediate schema (2), central persistance in elasticsearch database (3) and individual on-demand analysis with external tools (4). }
\label{fig:crawling-architecture}
\end{figure}

\begin{figure}[h!]
\centering
\includegraphics[width=\textwidth]{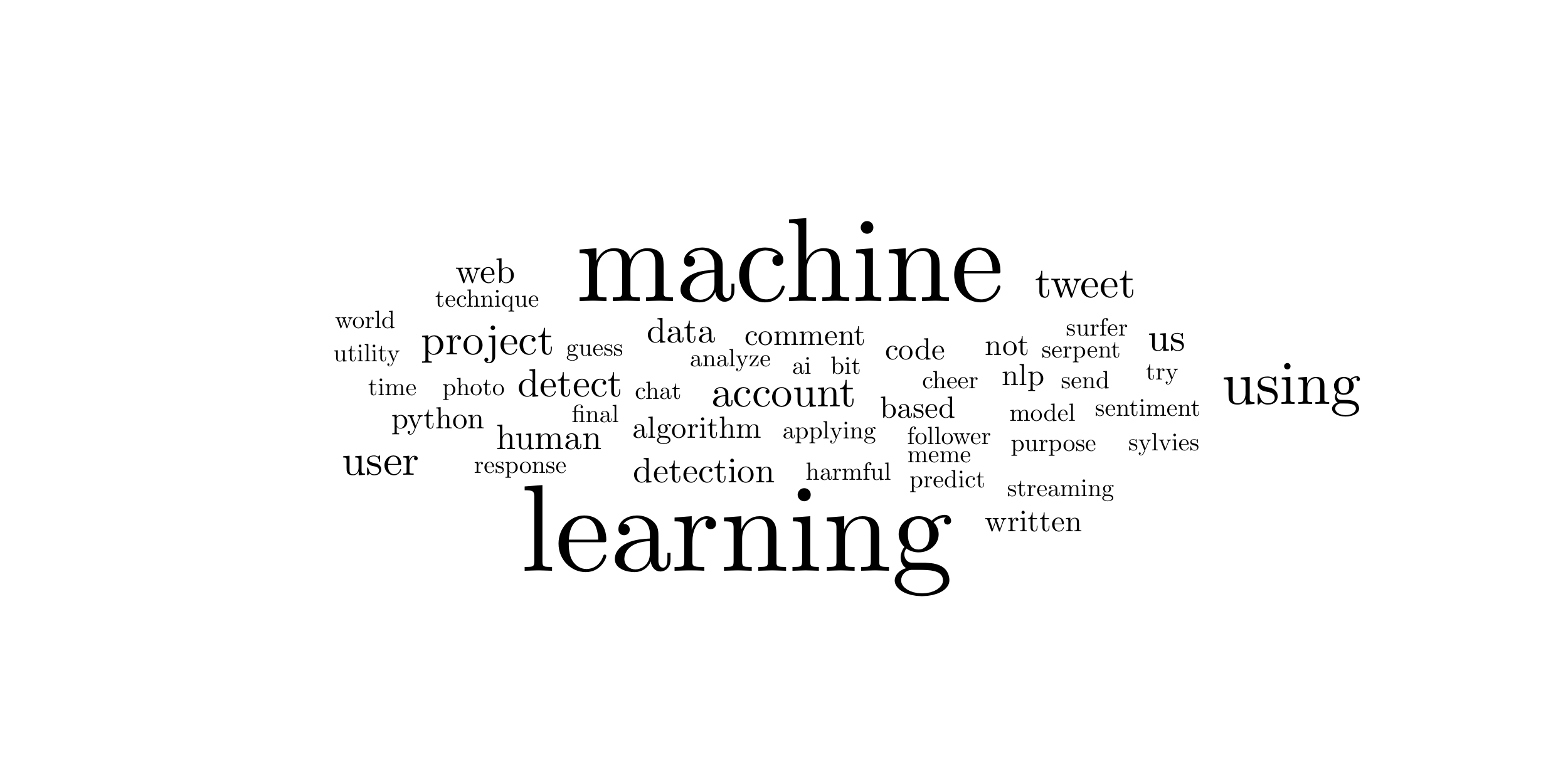}
\caption{Word cloud for repositories containing the search term "machine learning".}
\label{fig:wc-ml}
\end{figure}

\begin{algorithm}[h!]
\caption{Update procedure (see \cite{ER17Twitch}, we set $\lambda = 0.01$ and $r=0.06$)}
\label{Algorithm:update}
\begin{algorithmic}[1]
\Require $n_{\text{min}}$, $n_{\text{max}}$, $\lambda$, $t_{gap}$, $r$
\Initialize $t=0$, $MC=\emptyset$
	\While{stream is active}

	\State Read new text $x$ from stream \label{Algorithm:evoStream:read}
	\State Split $x$ into individual tokens
	\State Remove stopwords from tokens
	\State Build $n$-grams from the tokens $\forall n \in \{n_{\text{min}}, \ldots, n_{\text{max}}\}$
	

	\For{$i \gets 1, ..., \vert MC \vert$}
		\State $s_i \gets$ Compute cosine similarity between the vectors $mc_i[tf] \cdot idf$ and $c[tf] \cdot idf$
	\EndFor
		
	\State $j \gets \operatorname{arg\,min}_i(d_i)$ \Comment{find closest micro-cluster}
	
	\If{$s_j > r$}
		\State Merge $c$ into micro-cluster $mc_j$ \Comment{Merge into existing} \label{Algorithm:update:merge}
	\Else
		\State Add $c$ to set of clusters $MC$ \Comment{Establish new} \label{Algorithm:update:new}
	\EndIf

	\If{$t \mod t_{gap} = 0$}
		\State \Call{cleanup}{ } \label{Algorithm:update:cleanup}
	\EndIf
	
	\State $t \gets t + 1$
	
	\EndWhile
\end{algorithmic}
\end{algorithm}

\begin{algorithm}[h!]
\caption{Cleanup procedure}
\label{Algorithm:cleanup}
\begin{algorithmic}[1]
\Require $t_{gap}$, $MC$, $\lambda$, $r$
\Function{\textsc{cleanup}}{ }
	\ForEach{micro-cluster $mc \in MC$}
		\State $\Call{weight}{mc} \gets \Call{weight}{mc} \cdot 2^{-\lambda \Delta t}$ \Comment{fade micro-cluster}
		\If{$\Call{weight}{mc} \leq 2^{-\lambda t_{gap}}$}
			\State Remove $mc$ from the set of clusters $MC$
		\EndIf
		
		\ForEach{token $x \in mc$}
			\State \Call{weight}{$x$} $\gets$ \Call{weight}{$x$} $\cdot 2^{-\lambda \Delta t}$  \Comment{fade token}
			\If{\Call{weight}{$x$} $\leq 2^{-\lambda t_{gap}}$}
				\State Remove token $x$ from micro-cluster $mc$
			\EndIf
		\EndFor
	\EndFor

	\State Merge all $mc_i, mc_j$ where \Call{cosineSimiliarity}{$mc_i$, $mc_j$} $\leq r$

\EndFunction
\end{algorithmic}
\end{algorithm}

\end{document}